# Kerker Transform: Expanding Fields in a Discrete Basis of Directional Harmonics

Parker R. Wray‡ and Harry A. Atwater, †,*
‡Department of Electrical Engineering, California Institute of Technology, Pasadena, CA 91125, USA
†Thomas J. Watson Laboratories of Applied Physics, California Institute of Technology, Pasadena, CA 91125, USA
* corresponding haa@caltech.edu

## ABSTRACT

We present a linear coordinate transform to expand the solution of scattering and emission problems into a basis of forward and backward directional vector harmonics. The transform provides intuitive algebraic and geometric interpretations of systems with directional scattering/emission across a broad range of wavelength-to-size ratios. The Kerker, generalized Kerker, and transverse Kerker effect as well as other forms of highly directional scattering/emission are easily understood through open and closed loop contours in the complex plane. Furthermore, the theoretical maximum directivity of any scattering/emissive system is easily defined. The transformed far field harmonics have coordinates that are polar-angle invariant, interference between forward and backward harmonics weakly interact, and interference of same type harmonics alters directivity. Examples of highly directional scattering are presented including a Kerker scattering magnetic sphere, a directional scattering photonic nanojet, both under plane wave illumination, as well as generalized backward Kerker and transverse Kerker emission from sub-wavelength spheres that are near-field coupled to emitters. Solutions of scattering/emission under the Kerker transform are contrasted to the traditional Mie expansion for comparison.

## INTRODUCTION

The spherical harmonics are a set of fundamental modes of vibration on the sphere that provide valuable insights for understanding the scattering/emission of electromagnetic waves from particles[1–3]. Though these harmonics are invaluable to our understanding of scattering/emission from wavelength-sized objects, their atom-like spatial profile does not create a simple representation for describing phenomena such as angular momentum or directional scattering/emission. Fortunately, a linear transform of the spherical harmonics offers an insightful and mathematically simple method to study optical spin and orbital angular momentum[4,5]. In this manuscript we show that by a different but equally simple linear transform, the spherical harmonics can also give an intuitive basis to describe strongly directional scattering/emission. The resulting basis therefore provides a framework to study highly directional phenomena, which is of great value in many subjects in electromagnetics[6–16], while maintaining may of the beneficial properties that have popularized the spherical harmonics.

Under the Mie (vector spherical harmonic) framework, the Kerker effect is a method of achieving large forward-to-backward scattering/emission ratios through near exact cancelation of either the forward or backward intensity. This is achieved through the precise interference of same order electric ($\boldsymbol{\Psi}^{E}_{nmp}$) and magnetic-type ($\boldsymbol{\Psi}^{M}_{nmp}$) atom-like modes, where $n$ is the polar quantum number, $m$ is the azimuthal quantum number, and $p$ (0 = even and 1 = odd) is the azimuthal parity[16]. The case where generalized combinations of harmonics leads to (near) exact forward or backward cancelation is categorized as a generalized Kerker effect[17,13]. This requires no restriction on relative amplitude or phase between modes, only that they collectively interfere for exact cancelation in one direction. Cancelation in both directions is termed the transverse Kerker effect[12]. These conditions of cancelation in the exact forward or backward direction are formalized as the null points in the expression of exact forward ($\theta = 0$) or backward ($\theta = \pi$) power flow[18,19]. In principle, there are an infinite number of these null solutions. Besides a select set of simple examples (e.g., $c^{M}_{n1p} = c^{E}_{n1p}$), these solutions can be hard to intuit as they come from a quadratic polynomial of $n, m, p, t$ terms. Furthermore, directional scattering is a more holistic concept that need not evoke Kerker's conditions. It encompasses other metrics, such as directivity and side lobe behavior. The null conditions of the exact forward and backward power flow do not provide insight to these properties. Many works have shown that other combinations of interference not satisfying the null conditions (e.g., not evoking a Kerker effect) can give highly directional scattering/emission. For example, photonic nanojets typically have large number of side lobes with dominant and highly directive forward power flow, but backward power flow does not approach zero and can still be non-negligible[14]. Therefore, it is important to note that though the Kerker transform is named after the inspiring work of Milton Kerker, the Kerker basis is intended to efficiently represent all highly directional scattering /emission, not just the Kerker conditions. The basis is also intended to provide insights to metrics describing directionality such as directivity and side lobes.

To illustrate the general complexity of directional scattering in the Mie basis, consider two systems

$$\begin{aligned} &System\ 1: \quad -2a\boldsymbol{\Psi}_{111}^{E} - 2ib\boldsymbol{\Psi}_{211}^{E} + c\boldsymbol{\Psi}_{311}^{E} - ic\boldsymbol{\Psi}_{310}^{M} \\ &System\ 2: \quad -2a\boldsymbol{\Psi}_{111}^{E} - 2b\boldsymbol{\Psi}_{210}^{M} + c\boldsymbol{\Psi}_{311}^{E} - ic\boldsymbol{\Psi}_{310}^{M}, \end{aligned}$$

where $a, b, c \in \mathbb{Z}^{+} \ll \infty$. Are these systems directional? Which has the larger forward-to-backward ratio? What is occurring with respect to the side lobes? Can either system be Kerker, generalized Kerker, or transverse Kerker? What can we infer about directivity? How do the answers to these questions dependent on the choice of $a$, $b$, and $c$? These questions become more complex with the introduction of phase and as more harmonics are considered. The difficulty stems fundamentally from the Mie harmonics: they are designed to provide intuition of atom-like behavior, not directionality. To add further complication, many examples of directional scattering occur from particles which straddle the wave and ray-optic regime. E.g., photonic nanojets. When inclusions have around 2 appreciable harmonics (e.g., small sized inclusions), the conditions of harmonic interference giving rise to directionality is straightforward. In the limit of a very large number of harmonics (e.g., inclusions much larger than the wavelength) direct harmonic analysis is infeasible, and directionality is understood through a ray-optic approximation. Between the two regimes (2 – 50 harmonics) ray optics may not be accurate and wave optics not intuitive (e.g., the example of 4 harmonic systems proposed above).

The parameter space to achieve directional scattering/emission is vast. Inspired by the Kerker effect we propose a linear coordinate transform which seeks to provide a more intuitive basis for analyzing directional scattering/emission across all size regimes where wave optics is computationally viable, while still maintaining the useful properties which has made the spherical vector harmonics indispensable. The linear transform, termed the Kerker transform, and the resulting basis, termed the Kerker basis, is composed of forward and backward-type harmonics constructed from the Mie harmonics. The Kerker harmonics have the useful properties that forward and backward-type harmonics weakly interact with each other, and interference of same type harmonics is designed to control directivity and side lobes in the respective direction. The algebraic conditions for directional scattering under this framework is found to be simple to understand and have an intuitive geometric interpretation in the complex plane based on open and closed contours. Notably the conditions for Kerker scattering, transverse Kerker scattering (simultaneous suppression of both forward and backward intensity), and generalized Kerker scattering are easily conceptualized. The condition for theoretically maximal directivity is also easily conceptualized.

The difference between the Kerker and Mie harmonic expansions are summarized as:

*The Kerker framework easily represents directional scattering, while atom-like scattering arises from complicated interference. In the Mie framework, atom-like scattering is easily represented, while directional scattering arises from complicated interference.*

The remainder of this article is comprised of three sections. In the first section the Mie expansion is briefly reviewed and the Kerker expansion is presented. In the second section, features of the Kerker expansion studied in detail in the context of electromagnetic fields. Areas where the Kerker basis provides new beneficial insights for directional systems is emphasized, discussed, and contrasted to the Mie basis. The last section presents case studies of a Kerker, transverse Kerker, generalized Kerker, and general highly directional scattering/emitting systems, comparing the solutions in both the Kerker and Mie basis. Both scattering and emissive systems across a wide size-to-wavelength regime are discussed. We also provide an answer to our illustrative questions for *System 1* and *System 2*, posed above. Finally, we conclude with a summary of the results.

**DEFINING THE KERKER TRANSFORM**

Mie theory expands outward propagating scattered or emitted electromagnetic fields in terms of electric ($\boldsymbol{\Psi}_{nmp}^{E}$) and magnetic-type ($\boldsymbol{\Psi}_{nmp}^{M}$) spherical vector harmonics. As the names suggest, electric-type harmonics mimic electric atom-like multipole patterns in the far field, whereas magnetic-type harmonics mimic magnetic atom-like multipole field patterns. Time-harmonic electric and magnetic fields in the frequency domain are expanded under the Mie framework as

$$\begin{aligned} \boldsymbol{E} &= \sum_{n=1}^{\infty}\sum_{m=0}^{n}\sum_{p=0}^{1}\left(c_{nmp}^{M}\boldsymbol{\Psi}_{nmp}^{M}(\boldsymbol{r},k) + c_{nmp}^{E}\boldsymbol{\Psi}_{nmp}^{E}(\boldsymbol{r},k)\right) \\ \boldsymbol{H} &= \frac{-ik}{\mu\omega}\sum_{n=1}^{\infty}\sum_{m=0}^{n}\sum_{p=0}^{1}\left(c_{nmp}^{E}\boldsymbol{\Psi}_{nmp}^{M}(\boldsymbol{r},k) + c_{nmp}^{M}\boldsymbol{\Psi}_{nmp}^{E}(\boldsymbol{r},k)\right), \end{aligned} \tag{1}$$

where $c_{nmp}^{E}$ and $c_{nmp}^{M}$ are the complex electric and magnetic-type scattering coefficients, respectively. Though the coefficients are termed electric and magnetic-type, both coefficients actually scale either electric or magnetic-type harmonics depending on if one is viewing the electric or magnetic field. (E.g., the magnetic field scales the magnetic-type harmonic with the electric-type coefficient.) This is because $\boldsymbol{H} = \left(\frac{-i}{\mu\omega}\right) \nabla \times \boldsymbol{E}$ and the vector spherical harmonics change type under the curl operator: $\nabla \times \boldsymbol{\Psi}^t_{nmp}(\boldsymbol{r}, k) = k\boldsymbol{\Psi}^{1-t}_{nmp}(\boldsymbol{r}, k)$, where $t = 0 = M$ and $t = 1 = E$. The vector spherical harmonics are constructed from the scalar spherical harmonics and the spherical Bessel functions, as detailed in the appendix. All expansions in this text are written in a form general enough to represent any feasible electromagnetic field distribution in a linear, isotropic, and homogeneous host medium, with permeability, $\mu$, and permittivity, $\epsilon$. All fields are assumed time harmonic with angular frequency, $\omega$. Arbitrary time pulses are then generated though Fourier transformation. The harmonic time dependence is implied and not written explicitly. All bold variables are vectors in $\mathbb{C}^3$ or $\mathbb{R}^3$, dependent on the physical context, under the standard spherical basis $\hat{\boldsymbol{e}}_r$, $\hat{\boldsymbol{e}}_\theta$, and $\hat{\boldsymbol{e}}_\phi$. Spatial positions are denoted by $\boldsymbol{r} = \left(r\hat{\boldsymbol{e}}_r + \phi\hat{\boldsymbol{e}}_\phi + \theta\hat{\boldsymbol{e}}_\theta\right)$, where $r, \phi$, and $\theta$ are the radial, azimuthal, and polar coordinates, respectively. The wavenumber of the host media is $k^2 = \omega^2\epsilon\mu$. In both the Mie and Kerker expansions we adopt the convenient approach of assigning a type variable, $t \in [0,1]$, and a parity variable, $p \in [0,1]$, to write equations in compact form when applicable. Therefore, $1 - t$ is equivalent to flipping the harmonic type and $1 - p$ flips parity. This compact form helps to illuminate fundamental differences between the Mie and Kerker basis systems, which we feel is a critically important concept to convey as a first introduction to the Kerker transform. Correspondingly, we also use the cosine and sine expansion of the Mie harmonics (hence the parity variable and $m \geq 0$), because this also best illuminates' difference between the Mie and Kerker expansions. Though, we note that the complex azimuthal representation would elegantly simplify many of the analytic expressions discussed below. For this reason we encourage the reader to write the expansions on paper for each type and parity and also in the complex azimuthal form. Doing so will illuminate the simplicity of the Kerker transform, which is somewhat obscured by the chosen notation. Throughout the text, the summation bounds for $n$, $m$, $p$, and $t$ are the same as the bounds in equation 1. We will omit writing summation bounds explicitly and instead use the shorthand, $\Sigma_{nmpt}$.

The Kerker basis expands the electric and magnetic fields in terms of highly directional forward ($\boldsymbol{Y}^f_{nmp}$) and backward-type ($\boldsymbol{Y}^b_{nmp}$) harmonics. The field expansions under the Kerker framework are

$$\begin{aligned} \boldsymbol{E} &= \textstyle\sum_{nmp} \left(c^f_{nmp}\boldsymbol{Y}^f_{nmp}(\boldsymbol{r}, k) + c^b_{nmp}\boldsymbol{Y}^b_{nmp}(\boldsymbol{r}, k)\right) \\ \boldsymbol{H} &= \frac{k}{\mu\omega} \textstyle\sum_{nmp}(-1)^p \left(c^f_{nmp}\boldsymbol{Y}^f_{nm1-p}(\boldsymbol{r}, k) - c^b_{nmp}\boldsymbol{Y}^b_{nm1-p}(\boldsymbol{r}, k)\right), \end{aligned} \qquad 2$$

where $c^f_{nmp}$ and $c^b_{nmp}$ are the complex forward and backward scattering coefficients, respectively. Unlike the Mie expansion, the Kerker coefficients of one type multiply only the harmonics of the same type. (E.g., In both the electric and magnetic field, the forward coefficient always multiplies the forward harmonic.) This is because the Kerker harmonics do not alter type under the curl operation: $\nabla \times \boldsymbol{Y}^t_{nmp}(\boldsymbol{r}, k) = (-1)^{t-p} ik\boldsymbol{Y}^t_{nm1-p}(\boldsymbol{r}, k)$, where $t = 0 = f$ and $t = 1 = b$. Instead, the curl induces a parity change of the Keker harmonic, and changing parity is equivalent to an azimuthal phase shift, $\boldsymbol{Y}^t_{nmp}(r, \theta, \phi, k) = \boldsymbol{Y}^t_{nm1-p}\left(r, \theta, \phi + \frac{\pi}{2}, k\right)$. This type preservation under the curl operation is an important component to simplifying analytic expressions in directional systems. The Kerker harmonics are related to the vector spherical harmonics through the linear transform

$$\boldsymbol{Y}^t_{nmp}(\boldsymbol{r}, k) = (-1)^{t(n+m+1)}(i)^n \left(\boldsymbol{\Psi}^M_{nmp}(\boldsymbol{r}, k) + (-1)^{t-p} i\boldsymbol{\Psi}^E_{nm1-p}(\boldsymbol{r}, k)\right). \qquad 3$$

A complete component-wise expansion of the Kerker harmonics is shown in the appendix. The Kerker coefficients are related to the Mie coefficients through

$$\mathrm{c}^t_{nmp} = \frac{1}{2}(-1)^{t(n-m-1)}(-i)^n\left(c^M_{nmp} + (-1)^{1-t-p} ic^E_{nm1-p}\right), \qquad 4$$

where $t = 0 = f$ and $t = 1 = b$. Therefore, electromagnetic scattering/emission problems can be solved in either the Kerker or Mie basis, then subsequently transformed into the other basis when beneficial for analysis.

It is important to note that though the Kerker harmonics are formed from a superposition of electric and magnetic-type Mie harmonics, this basis is not equivalent to the transform used to study angular momenta. The Kerker basis does not preserve circular polarization or helicity (e.g., it is not equivalent to $\boldsymbol{\Psi}^M_{nmp} \pm \boldsymbol{\Psi}^E_{nmp}$). The Kerker harmonics are not eigenvectors of the orbital angular momentum operator ($\frac{1}{k}\nabla \times$), as evident by the change in parity under the curl discussed above. With that said, multiple reports have studied the connection between Kerker's conditions and helicity preservation that exists in suitably rotationally symmetric scattering/emissive systems[5,20]. Such connections can also be studied in the Kerker basis by forming Kerker harmonics that also preserve handedness. This is achieved through a linear transform of the regular Kerker harmonics to incorporate both parities. E.g., $\boldsymbol{Q}^t_{nmh}(\boldsymbol{r}, k) = \boldsymbol{Y}^t_{nm0}(\boldsymbol{r}, k) + (-1)^{t-h} i\boldsymbol{Y}^t_{nm1}(\boldsymbol{r}, k)$, where $h = 0 = L$ for left-handed polarization and $h = 1 = R$ for right-handed polarization. This basis is an eigenvector of the angular momentum operator as, $\nabla \times \boldsymbol{Q}^t_{nmh} = (-1)^h k \boldsymbol{Q}^t_{nmh}$. Therefore, the Kerker harmonics can be used to efficiently understand directional scattering in systems with and without angular momentum preservation.

## FEATURES OF THE KERKER TRANSFORM

A primary benefit of the Kerker basis is having a simplified expression in the far field compared to the Mie harmonics. The far field Kerker harmonics are

$$\mathbf{Y}^t_{far,nmp}(\boldsymbol{r},k) = i\frac{e^{ikr}}{kr} X^t_{nm}(\theta) \begin{bmatrix} 0\ \hat{\boldsymbol{e}}_r \\ +sin\left(m\phi - p\frac{\pi}{2} + t\pi\right)\hat{\boldsymbol{e}}_\theta \\ +cos\left(m\phi - p\frac{\pi}{2}\right)\hat{\boldsymbol{e}}_\phi \end{bmatrix} + O\left\{\frac{1}{(kr)^2}\right\}, \quad 5$$

where

$$X^t_{nm}(\theta) = \frac{(-1)^{t(n+m+1)}}{\sqrt{2n(n+1)}}\left(\tau^m_n(\theta) + (-1)^t m\pi^m_n(\theta)\right) \quad 6$$

is a real valued function describing the forward, $X^f_{nm}(\theta)$, and backward, $X^b_{nm}(\theta)$, polar-angle dependence. Equation 5 shows that the vector components differ only by simple trigonometric relations. Equation 6 defines the relationship of the Kerker polar angle functions to the Mie polar angle functions, $\tau^m_n(\theta)$ and $m\pi^m_n(\theta)$. By construction, all vector components of the far field Kerker harmonics share the same polar-angle dependence. This polar-angle invariance is a key feature of the Kerker harmonics and allows us to focus on $X^t_{nm}$ in order to understand directional properties. In contrast, the Mie expansion has a different polar-angle expansion for each vector component.

Figure 1 plots both the Kerker ($X^t_{nm}$) and Mie ($\tau^m_n, m\pi^m_n$, upper quadrant) polar angle functions up to the quantum numbers $n = 4$ and $m = 4$.

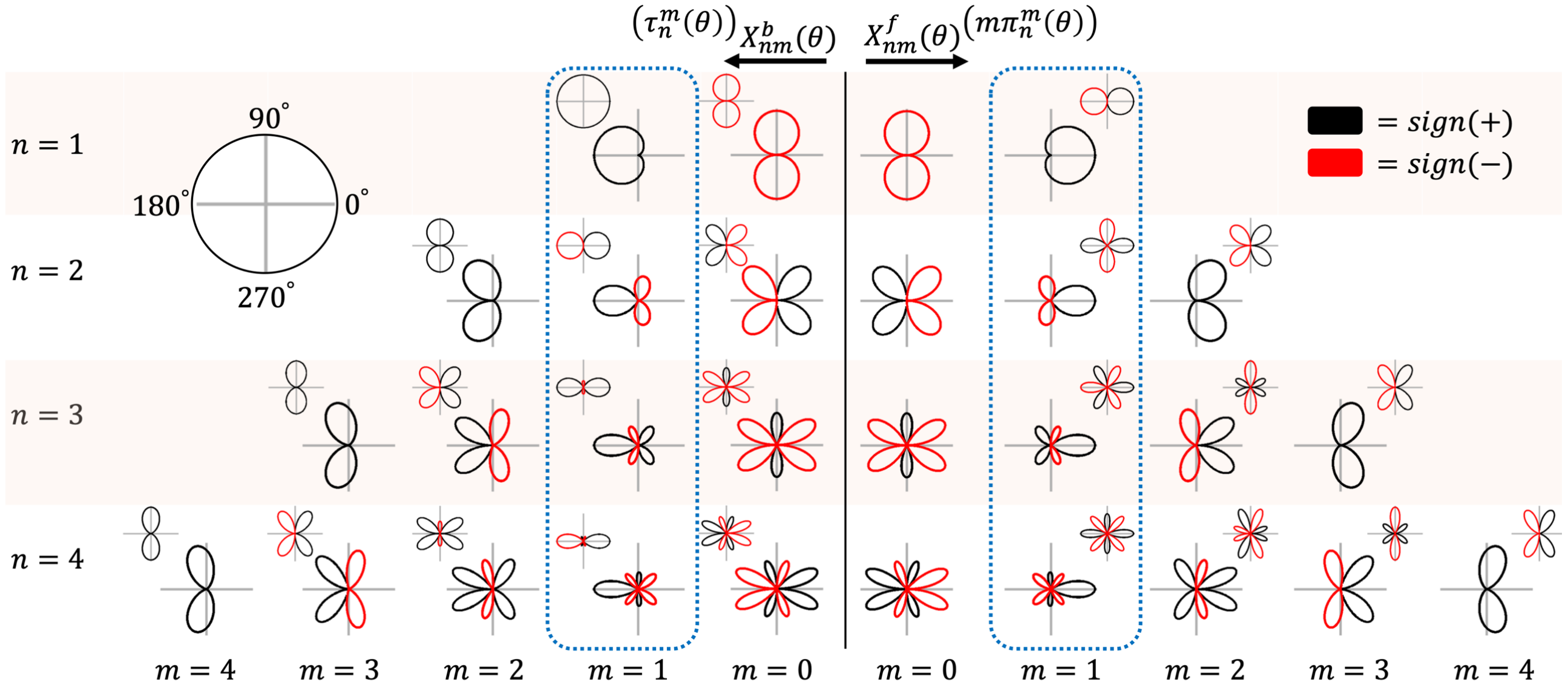

**Figure 1.** Table of the Kerker and Mie (upper corner) polar angle functions. Rows and columns designate polar and azimuthal quantum numbers, respectively. Forward and backward-type functions and corresponding $m\pi^m_n$ and $\tau^m_n$ functions are on the right and left-hand side, respectively. The functions are plotted in polar coordinates where the polar angle is given by the key in the upper left. Positive and negative values of the radius are denoted by black and red lines, respectively. The columns encircled by dashed blue lines contain the polar functions that have nonzero values in the exact forward or backward direction.

From this figure we note three features of $X^t_{nm}$, designed for convenience when studying directional systems:

(1) $X^t_{nm}$ has a simple relation between the two types: the backward Kerker functions are the forward Kerker functions rotated 180°. (I.e., $X^t_{nm}(\theta) = X^{1-t}_{nm}(\pi - \theta)$.) This is not just convenient to conceptualize, but also allows only one type to be calculated and stored in memory. In contrast, the Mie polar angle functions are not related through a rotation and have completely different shapes. This is because $\pi^m_n$ is related to the associated Legendre polynomial and $\tau^m_n$ is related to the derivative of the associated Legendre polynomial. Furthermore, 180° rotations of the Mie polar functions lead to different inversion parity

relations that are dependent on the quantum polar numbers. I.e., $\tau_n^m(\pi-\theta)=(-1)^{n-m+1}\tau_n^m(\theta)$ and $\pi_n^m(\pi-\theta)=(-1)^{n-m}\pi_n^m(\theta)$. This further complicates deriving intuitive interference relations for the Mie harmonics because the sign of the lobes alters as a function of polar number and harmonic type. The Kerker functions need no such sign relation.

(2) The Kerker polar functions are highly directional and have the clear notion of primary and side lobes. Therefore, the functions can easily represent directional fields. This is unlike the Mie counterpart, where there is no clear definition of side lobes. For each Kerker harmonic, the total number of lobes is given simply by $n-m+1$ and the proportion of the side lobes in the nondominant hemisphere is given by $ceil\left(\frac{n-m}{2}\right)$.[§1] Furthermore, harmonics with larger polar quantum numbers have narrower beam widths for all lobes. Therefore, from knowing just the quantum numbers of a Kerker harmonic you can infer far field directionality, relative beam widths, and side lobes, including the side lobe concentration in both the forward and backward hemispheres for each harmonic. The Mie polar functions provide no such intuition, as they are not designed for this purpose. For example, contrast the field profiles of $\tau_3^1$ and $\pi_3^1$. The number of lobes, amplitude of lobes, width of lobes, and sign of the lobes from these Mie polar functions are all different.

(3) The Mie polar functions are neither mutually orthogonal nor orthogonal to each other. In contrast, the Kerker polar functions of the same type and azimuthal number are orthogonal over the domain $\int_0^\pi \partial\theta \sin(\theta) X_{nm}^t(\theta) X_{n'm}^t(\theta)$.

Like Mie theory, the $m=1$ column in figure 1 (circled in dashed blue) is the only column to have a nonzero exact forward or backward field. This column is particularly important and usually predominant on physical grounds. For example, symmetries of the scattering/emitting object, such as being of spherical shape, can preclude $m\neq 1$. In the Kerker basis the primary lobes of the $m=1$ column are exactly centered in either the forward ($\theta=0$) or backward ($\theta=\pi$) direction, with exactly no field in the opposite direction. We will show later this attribute simplifies the analytic expression for calculating forward-to-backward ratios. With that said, the Kerker basis is a directional expansion for all $m\geq 1$, as evident in figure 1. It can describe any arbitrary scattering/emissive system as an expansion of directional harmonics, given that the fields can be represented by a spherical harmonic expansion.[§2]

From the surface equivalence principle, the scattered/emitted field can also be represented as electric ($\boldsymbol{J}=\hat{\boldsymbol{e}}_r\times\boldsymbol{E}$) and magnetic ($\boldsymbol{M}=-\hat{\boldsymbol{e}}_r\times\boldsymbol{H}$) current densities. Such relations are of interest in applications such as near-to-far transformation. The Kerker basis is paired to a corresponding basis of forward ($\boldsymbol{\jmath}_{nmp}^f$) and backward-driving ($\boldsymbol{\jmath}_{nmp}^b$) current densities, where $\boldsymbol{\jmath}_{nmp}^t(\boldsymbol{r},k)=\hat{\boldsymbol{e}}_r\times\boldsymbol{\Upsilon}_{nmp}^t(\boldsymbol{r},k)$. Given that

$$\boldsymbol{\jmath}_{far,nmp}^t(\boldsymbol{r},k)=i\frac{e^{ikr}}{kr}X_{nm}^t(\theta)\begin{bmatrix}0\,\hat{\boldsymbol{e}}_r\\ -cos\left(m\phi-p\frac{\pi}{2}\right)\hat{\boldsymbol{e}}_\theta\\ sin\left(m\phi-p\frac{\pi}{2}+t\pi\right)\hat{\boldsymbol{e}}_\phi\end{bmatrix}+O\left\{\frac{1}{(kr)^2}\right\}, \qquad 7$$

we find that all beneficial properties of the Kerker far field harmonics also equally apply to the Kerker far field current densities. In particular, the $X_{nm}^t$ dependence is unchanged. This provides a method to understand what current distributions give rise to directional scattering in the far field.

Using the far field Kerker basis, the time-averaged far field Poynting vector is

$$\langle\boldsymbol{S}_{far}\rangle=\frac{1}{2}\frac{1}{\mu\omega kr^2}\left(\|A(\phi,\theta)\|^2+\|B(\phi,\theta)\|^2\right)\hat{\boldsymbol{e}}_r, \qquad 8$$

where

$$\begin{aligned}A(\theta,\phi)&=\sum_{nmp} cos\left(m\phi-p\frac{\pi}{2}\right)\left(c_{nmp}^f X_{nm}^f(\theta)+c_{nmp}^b X_{nm}^b(\theta)\right)\\ B(\theta,\phi)&=\sum_{nmp} sin\left(m\phi-p\frac{\pi}{2}\right)\left(c_{nmp}^f X_{nm}^f(\theta)-c_{nmp}^b X_{nm}^b(\theta)\right).\end{aligned} \qquad 9$$

---

[§1] Note that the domain of the polar angle functions is $\theta\in[0,\pi]$ and lobes are counted within this angular region.

[§2] Just like the Mie harmonics, the Kerker harmonics are defined with respect to a global coordinate system. To efficiently represent a directional beam that is off-axis to the globally defined forward/backward direction, either the system can be solved under a new global coordinate system that aligns with the direction of interest or the rotation theorem for vector spherical harmonics, based on the Wigner-D functions, can be used.

The $\left(c^f_{nmp}X^f_{nm}(\theta) \pm c^b_{nmp}X^b_{nm}(\theta)\right)$ terms in equation 9 shows that it is instructive to understand how the forward and backward polar angle functions interfere with each other. Luckily, the polar angle functions are designed to concentrate energy in their respective dominant hemisphere. Therefore, interference between forward and backward harmonics is weak. Alternatively stated, primary lobes of one type (forward/backward) interact only with side lobes of the other type (backward/forward). Figure 2 illustrates this concept. From this figure we see that weak interaction enables a convenient method to intuit the interference between modes of different type. They can be viewed as *approximately* noninteracting in their respective dominant hemisphere. §3 This provides a rule-of-thumb for approximating harmonic interference in complicated systems. In contrast, the Mie harmonics have strong interreference between the electric and magnetic-types and the resulting scattering/emission lobes have no rule-of-thumb behavior.

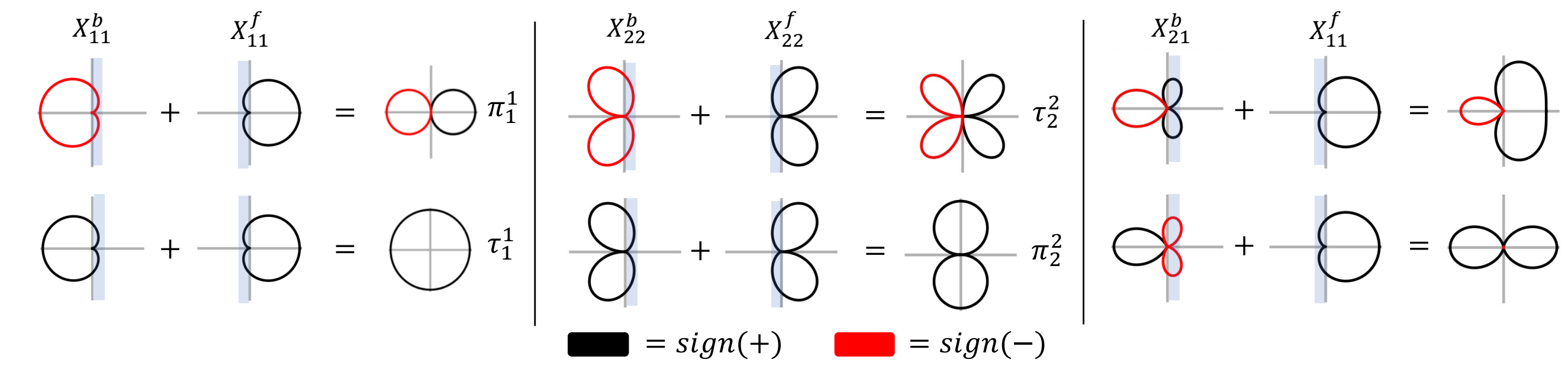

**Figure 2.** Examples of interference between opposite type Kerker polar angle functions. The color convention and angle orientation follow the definition from figure 1. Therefore, the top and bottom row correspond to $X^f_{nm} - X^b_{nm}$ and $X^f_{nm} + X^b_{nm}$ interference, respectively. The shaded region highlights the non-dominant hemisphere for each function. The left and middle example show how the Mie functions can be recovered, while the right example is a more general interference between different polar quantum numbers. The functions are plotted in polar coordinates where the polar angle is given by the key in the upper left. Positive and negative values of the radius are denoted by black and red lines, respectively.

The right most example in figure 2 shows the most general form of interference between polar angle functions of different type and quantum numbers. The left and center examples in figure 2 show interference of opposite type polar angle functions with the same quantum numbers. These two cases are important because they represent the inverse transform that recovers the Mie angular functions. This result can be seen by rearranging equation 6 to show that $X^f_{nm}(\theta) + (-1)^{n+m+1}X^b_{nm}(\theta) = 2\tau^{|m|}_n(cos\,\theta)$ and $X^f_{nm}(\theta) - (-1)^{n+m+1}X^b_{nm}(\theta) = 2m\pi^{|m|}_n(\cos\theta)$. More generally, atom-like fields are achieved in the Kerker basis through interference that gives rise to the inverse Kerker transform:

$$\mathbf{\Psi}^{\mathrm{t}}_{nmp}(\boldsymbol{r}) = (-1)^{t(1-p)} i^{-(n+t)}\,\frac{1}{2}\,\boldsymbol{\Upsilon}^f_{nmt-p}(\boldsymbol{r}) - (-1)^{n+m+t}\boldsymbol{\Upsilon}^b_{nmt-p}(\boldsymbol{r}) \qquad 10$$

where, again, $\mathbf{\Psi}^{\mathrm{t}}_{nmp}(\boldsymbol{r})$ are the Mie vector harmonics with $t = 0 = M$ and $t = 1 = E$. Equation 10 and equation 3 formalize our italicized summary in the introduction. Directional fields require complicated interference in the Mie expansion and atom-like fields require complicated interference in the Kerker expansion.

From the far field Poynting vector, the far field intensity is defined as $I(\theta,\phi) = \langle \boldsymbol{S}_{far}\rangle \cdot r^2\hat{\boldsymbol{e}}_r$. Integrating this intensity over the azimuthal direction gives

$$I(\theta) = I^f_0 + I^f_1 + I^b_0 + I^b_1 = \frac{\pi}{\mu\omega k}\sum_{tmp}(1+\delta_{m0})\left\|\sum_n c^t_{nmp}\,X^t_{nm}(\theta)\right\|^2, \qquad 11$$

where the $(1+\delta_{m0})$ term comes from the fact that $c^M_{n01} = c^E_{n01} = 0$. Interestingly equation 11 shows that the azimuthally integrated intensity is truly not dependent on interference between the forward and backward harmonics or interference between different parity. This is, again, another useful feature of the Kerker harmonics. The total azimuthally integrated intensity can be viewed as resulting from four noninteracting partial fields, each with intensity $I^t_p(\theta) = \frac{\pi}{\mu\omega k}\sum_m(1+\delta_{m0})\left\|\sum_n c^t_{nmp}\,X^t_{nm}(\theta)\right\|^2$. For any given polar angle, these partial intensities have the geometric interpretation as being the sum of the distances from the origin occurring from the tip-to-tail coherent addition of $\sum_n c^t_{nmp}X^t_{nm}(\theta)$. This result allows for an intuitive geometric interpretation of directional scattering, which will become more evident later in this section. It is also worth noting that four partial fields represent the most general form. In systems with symmetry, such as plane wave or dipole excitation of a sphere, one forward and one backward partial field completely describes the system. Furthermore, it is often the case that only the $m = 1$ terms are appreciable. Therefore, it is common that multiple simplifications to equation 11 are applicable.

§3 Since side lobes are concentrated closer to the horizontal ($\theta = \pi/2$), interactions between forward and backward harmonics are more pronounced near this angular region.

Equations 8 and 11 highlight the importance of understanding interference between Kerker polar angle functions of the same type. Figure 3 defines this relationship for the forward polar angle functions. We omit examples of the backward functions because, unlike the Mie harmonics, the inversion symmetry implies the results are the same just rotated 180°. Figure 3 shows that constructive interference of same type harmonics results in an increased primary lobe and an overall more directive far field. Likewise, destructive interference decreases the primary lobe and reduces directivity. This provides an intuitive interference relationship to identify directive systems. Adding coefficients of the same type increases directivity. Subtraction reduces directivity. This condition can be easily generalized to complex valued coefficients giving an intuitive geometric interpretation based on coefficients as vectors in the complex plane. From this picture, same type Kerker coefficients pointing in a similar direction will increase directivity. Coefficients pointing in opposite directions will decrease the directivity.

In the exact forward and backward directions, the Kerker polar angle functions are designed to take the simple form

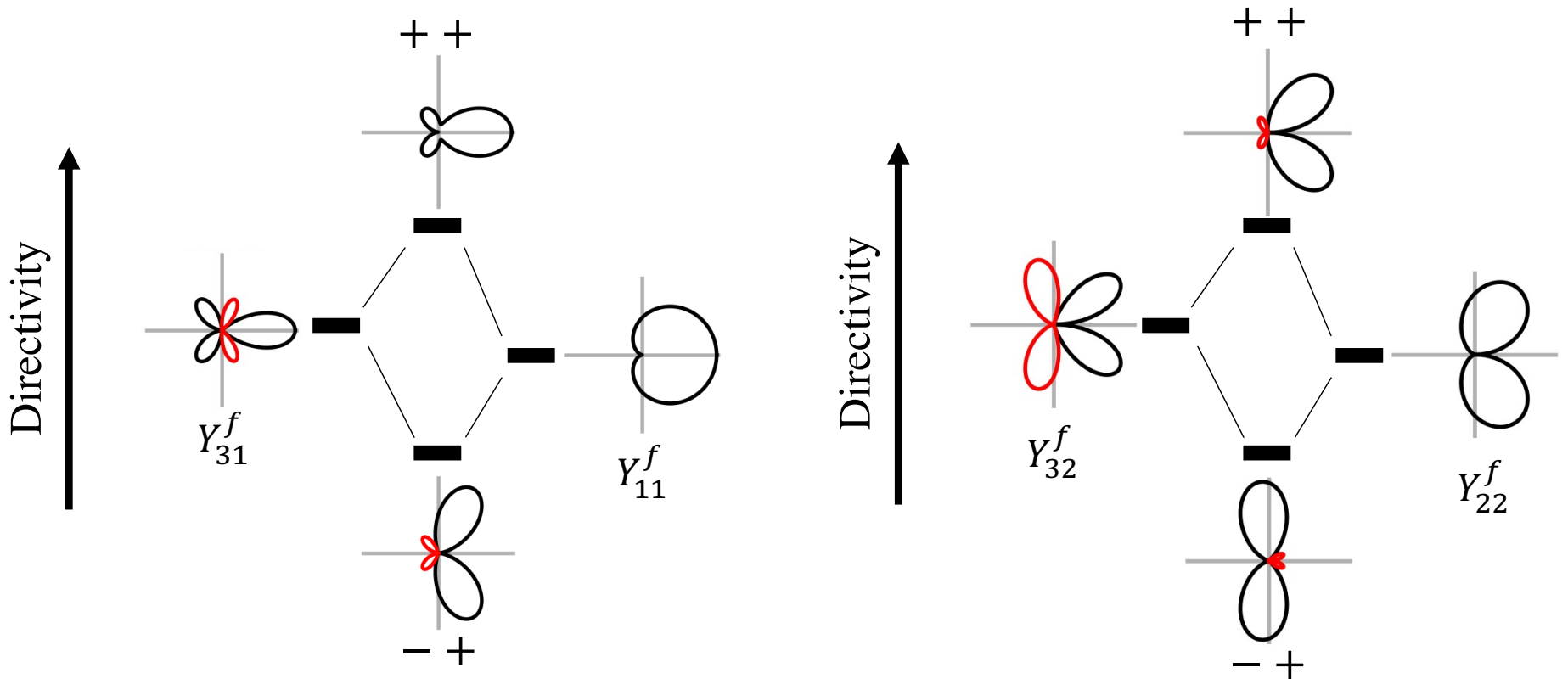

**Figure 3.** Examples of interference between same type Kerker polar angle functions. The left example is a combination of same parity polar numbers when $m = 1$. The right example is a combination of opposite parity polar numbers for azimuthal numbers that do not have exact forward scattering/emission ($m \neq 1$).

$$X_{nm}^f(\theta = 0) = \tfrac{1}{2} K_n \delta_{m1} \quad X_{nm}^f(\theta = \pi) = 0$$
$$X_{nm}^b(\theta = 0) = 0 \qquad X_{nm}^b(\theta = \pi) = \tfrac{1}{2} K_n \delta_{m1}, \qquad 12$$

where $K_n = \sqrt{(2n+1)}$. Equation 12 formalizes a property of the Kerker harmonics that can be inferred from figure 1. The Kerker harmonics have an exact forward or backward lobe only for the $m = 1$ quantum number. Furthermore, these functions have exactly zero field in the opposite direction. Therefore, there is always complete noninteraction between forward and backward harmonics in the exact forward and backward directions. This property enables a simplified and geometrically intuitive expression for calculating exact forward and backward intensities and forward-to-backward ratios.

The far field intensity in the exact forward and backward directions is

$$I(\theta = 0) = \frac{\pi}{2\mu k \omega}\left(\sum_p \left\| \sum_n K_n c_{n1p}^f \right\|^2\right)$$
$$I(\theta = \pi) = \frac{\pi}{2\mu k \omega}\left(\sum_p \left\| \sum_n K_n c_{n1p}^b \right\|^2\right). \qquad 13$$

Equation 13 shows the exact forward and backward intensity can be understood geometrically as the magnitude of the vector that results from coherently adding scaled forward and backward coefficients in the complex plane, $\sum_n K_n c_{n1p}^t$. The forward-to-backward ratio is then the ratio of the lengths of these vectors. This provides a useful geometric connection between the Kerker coefficients and the resulting forward and backward intensity. When vectors added together approach a closed loop, there is weak scattering/emission in that direction. Equation 13 is a specific example of equation 11, for the important case where $\theta = 0$ or $\pi$.[§4] Under this condition, the scaling factor $X_{nm}^{t}$ takes the simplified form given by equation 12. For an arbitrary direction, the same vector addition rules apply but the scaling factors are based on Equation 11.

Equation 13 provides intuitive geometric conditions to understand the Kerker effects. Forward or backward Kerker scattering can now be viewed as the special case when either all $c_{n1p}^{b}$'s or $c_{n1p}^{f}$'s are zero, respectively. E.g., a forward Kerker scattering object will have no backward Kerker coefficients, $c_{n1p}^{b}$. This property is why the coefficients are termed "Kerker coefficients". Generalized forward or backward Kerker scattering can also be understood as occurring when either $\sum_n K_n c_{n1p}^{b}$ or $\sum_n K_n c_{n1p}^{f}$ are zero, respectively. This corresponds to vectors of one type that, when added head-to-tail, form a closed loop in the complex plane. The transverse Kerker effect occurs when both vectors of both types form a closed loop. I.e., $\sum_n K_n c_{n1p}^{b}$ and $\sum_n K_n c_{n1p}^{f}$ are zero. Forms of directional scattering which do not obtain identically zero forward or backward fields are identified by comparing the length of the coherently added forward vectors versus the backward vectors. I.e., the forward-to-backward ratio is then the ratio of the length to the total forward to the total backward vectors. Note that these conditions apply for all relevant parities used to describe the field. Figure 4 gives a schematic of the geometric representations of different types of directional scattering based on the Kerker coefficients. These are substantially easier interpretations compared to the complex modal interference relationships necessary to satisfy these conditions in the Mie framework. This will be further discussed through examples in the next section.

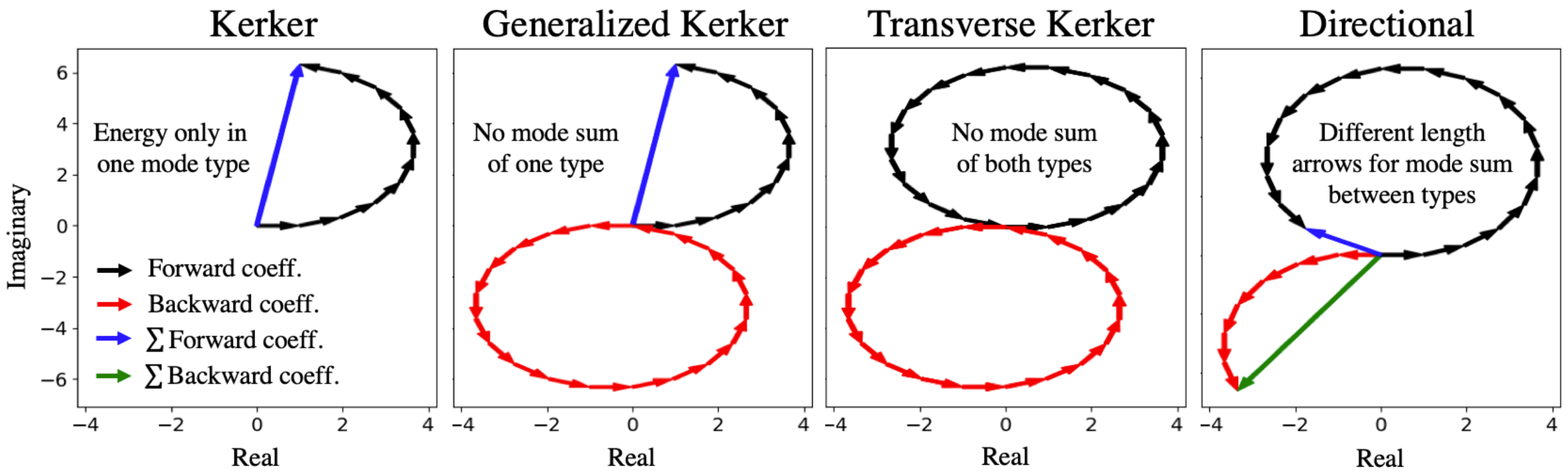

**Figure 4.** Schematics of different types of directional scattering/emission represented as closed and open-loop paths in the complex plane. Individual forward and backward modes are given by black and red arrows, respectively. Modes of the same type are connected head-to-tail and progressively increment from $n = 1$ (tail at the origin) to $n = n_{max}$. The coherent sum of the forward and backward modes is designated by blue and green arrows, respectively. These arrows start at the origin and connect to the tip of the max polar number vector. The left most schematic is an example of forward Kerker behavior, where no backward modes are present. The left middle example is a forward generalized Kerker effect where backward modes coherently cancel in the exact backward direction. The right middle example shows the transverse Kerker effect where both modes coherently cancel the exact forward and backward direction leaving only transverse (side lobe) scattering/emission. The right most schematic is an example of general backward preferential scattering/emission.

Finally, it is instructive to consider the expression for total power flow and directivity under the Kerker expansion. Like the vector spherical harmonics, the Kerker harmonics are orthogonal on the sphere. Therefore, the total scattered/emitted power is

$$W_{scatter/emit} = W_0^f + W_1^f + W_0^b + W_1^b = \frac{\pi}{2\mu\omega k}\sum_{tnmp}(1+\delta_{m0})\left\|c_{nmp}^{t}\right\|^2 , \quad 14$$

where the total power is composed of the forward or backward partial powers, $W_p^t = \frac{\pi}{2\mu\omega k}\sum_{nm}(1+\delta_{m0})\left\|c_{nmp}^{t}\right\|^2$. Unlike the Mie harmonics which distribute the total power into electric and magnetic-type excitations, equation 14 shows the Kerker harmonics distribute power into forward and backward-type excitations. This helps give intuition on the fraction of the total power concentrated into a particular hemisphere. This fraction can also be understood geometrically, where each partial power (and therefore the total power) is given by the arclength of the scattering coefficients added head-to-tail in the complex

[§4] Like equation 11, equation 13 is written in the most general form for any arbitrary system. If the scattering/emissive system has the proper symmetry, equation 13 can be simplified such that only one parity for each type is necessary. Other simplifications such as reduced azimuthal or polar orders can also apply.

plane. I.e., A longer arclength means a larger proportion of the total power is concentrated into that harmonic type. Dividing the origin-to-tip vector length of equation 13 by the arclength of equation 14 then gives an intuitive definition of forward or backward directivity as

$$D^{t'} = 4\pi \frac{\sum_p \left\| \sum_n K_n c^{t'}_{n1p} \right\|^2}{\sum_{tnmp} (1+\delta_{m0}) \left\| c^t_{nmp} \right\|^2}, \qquad 15$$

where $D^f = D(\theta = 0)$ and $D^b = D(\theta = \pi)$. Equation 15 formalizes the argument of directivity presented in figure 3 and directly connects directivity to the behavior of Kerker coefficients in the complex plane. Directivity is proportional to origin-to-tip length and inversely proportional to arclength. From this framework we can rigorously derive the conditions to maximize directivity and relate these conditions to intuitive curves in the complex plane.

As more complex coefficients of the same type (each represented as a vector in the complex plane) point in a similar direction in the complex plane the numerator of equation 15 increases while the arclength remains unchanged. The triangle inequality enforces that the numerator of equation 15 is maximized when all vectors of the same type point in the exact same direction, $\sum_p \left\| \sum_n K_n c^{t'}_{n1p} \right\|^2 = \sum_{np} \left\| K_n c^{t'}_{n1p} \right\|^2$. I.e., the curve formed by head-to-tail addition of the coefficients forms a straight line. The geometric representation of all vectors pointing in the same direction is the condition of perfect constructive interference. Though, to maximize directivity, the denominator of equation 15 should also be minimized. To achieve this, all coefficients in the denominator that are not present in the numerator should be zero. Since $K_n > 1$, we can conclude that:

*The theoretically maximal directivity for a system with $n_{max}$ harmonic orders occurs when you satisfy Kerker's condition and all Kerker coefficients constructively interfere.*

Therefore, Kerker scattering is a necessary but not sufficient condition to achieve the theoretical maximum directivity. Furthermore, generalized Kerker can never achieve the theoretically maximal directivity because though the origin-to-tip length in the unwanted direction is zero, the arclength for that direction is not zero.

## EXAMPLES

To highlight the usefulness of the Kerker transform, we give four instructive examples of directional scattering and study their results under the Kerker and Mie expansions. The goal in this section is to provide examples of when it can be useful to switch from the Mie to the Kerker framework. In order to highlight the generality of the Kerker expansion, we study both nearfield and far-field excitations of sub-wavelength and larger than wavelength particles. All examples are summarized in Figure 5.

The first and second row of figure 5 shows a schematic of each system and their corresponding azimuthally integrated far field polar intensity profile, respectively. The left example is the scattering response of a sphere with the material properties initially proposed by Milton Kerker to explain Kerker scattering; the case where $\epsilon = \mu$. The response has exactly no backward field and Kerker's condition is satisfied for all quantum numbers supported by the sphere. The middle-left example is of generalized backward Kerker emission, where near complete suppression of the forward intensity is achieved. The system achieving this emission is composed of the combined response from a 374nm wavelength emitter near-field coupled 90nm below a 164nm $TiO_2$ sphere ($\eta = \sqrt{\epsilon_r \mu_r} = 2.42$)[21]. The middle-right example is of transverse Kerker scattering from a 120nm Si sphere ($\eta = 3.92 + i2.49E^{-2}$)[22], achieved by coupling two 609nm wavelength emitters to the sphere. One emitter is located 204nm above and below the sphere, respectively. The right most example is of highly directional forward scattering by creating a photonic nanojet. This scattering is achieved by illuminating a 1200nm $SiO_2$ sphere ($\eta = 1.43 + i2.52E^{-3}$)[23] with a 400nm plane wave. In all cases, the background medium is assumed to be air. The solution to the scattering by a sphere illuminated by a plane wave or a dipole emitter can be found in [19] and [24], respectively.

The third and fourth row of figure 5 plots the $K_n$-scaled Kerker (upper row) and Mie (lower row) coefficients, respectively, as vectors in the complex plane. This plotting method is commonly used as it describes both amplitude and phase, which is necessary to understand directional scattering [8,25]. The left example clearly shows forward Kerker scattering as the backward coefficients (red arrows) satisfy the Kerker condition that all $c^b_{n10}$'s are zero. The forward coefficients (black arrows) constructively interfere leading to a nonzero total forward intensity (blue arrow). Alternatively, determining Kerker's forward condition using the Mie coefficients requires a systematic comparison of both the angle and phase relationship between each pair of electric and magnetic-type harmonics. Though this is a tractable task for $n_{max} \approx 3$, it is still hard to say for certain that the system is exactly satisfying the Kerker forward condition without using a ruler and protractor.

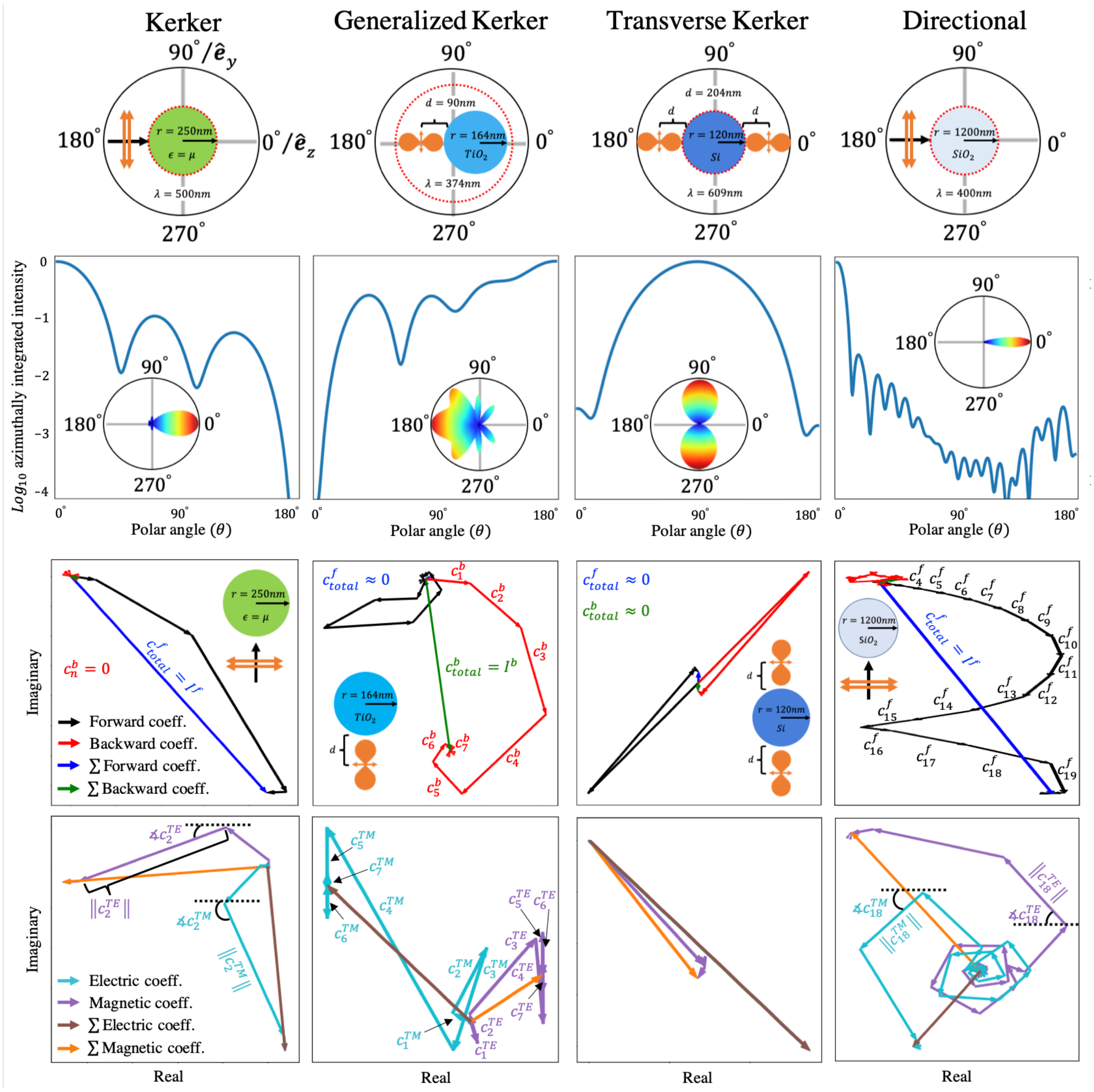

**Figure 5.** Schematics of highly directional scattering/emission (row 1), corresponding log base 10 azimuthally integrated far field with polar intensity plots (row 2), and corresponding $K_n$-scaled Kerker (row 3) and Mie (row 4) coefficients as vectors in the complex plane. The first example (column 1) is of an exact Kerker scattering system composed of a 250nm radius magnetic sphere ($\epsilon = \mu$) excited by a 500nm wavelength plane wave traveling in the $\hat{\boldsymbol{e}}_z$ direction. The second example (column 2) shows generalized backward Kerker emission achieved by near field coupling a 164nm $TiO_2$ sphere to a dipole emitting at 374nm. The dipole is located 90nm below the bottom of the sphere on the z-axis and has a moment in the $\hat{\boldsymbol{e}}_y$ direction. The third example (column 3) is of transverse Kerker scattering achieved in a 120nm Si sphere excited simultaneously by two dipoles, both emitting at 609nm. The dipoles are 204nm above and below the sphere, respectively. Both dipoles are on the z-axis with moments in the $\hat{\boldsymbol{e}}_y$ direction. The final example (column 4) is of highly directional scattering from a photonic nanojet made from a 1200nm $SiO_2$ sphere excited by a plane wave with a 426nm wavelength. In all cases the sphere is centered at the origin and the region outside of the red dashed circle defines the domain of validity for the expansion. All intensity is normalized as $I(\theta)/\max(I)$ and share the same log scale. Coefficients of the same type are connected head-to-tail and progressively increment from $n = 1$ (tail at the origin) to $n =$

The Kerker coefficients in the middle-left example show generalized backward Kerker behavior. This is evident by the coherent sum of the forward coefficients forming a closed loop, $\sum_n K_n c_{n10}^f \approx 0$, and the coherent sum of the backward coefficients producing a nonzero open loop for the total backward intensity (green arrow). The arc in the path length of the backward coefficients as well as the loop of the forward coefficients indicate the presence of excess side lobes since the vectors are not strictly in the same direction. These side lobes are evident in the azimuthally integrated intensity. The Mie coefficients traverse a sporadic pattern in the complex plane. With $n_{max} \approx 7$ and no easily discernable interference relationship, these coefficients do not illuminate directional emission or properties of side lobes. Clearly the Mie coefficients are not the appropriate tool for this problem.

The middle-right example shows transverse Kerker behavior as evident by both the forward and backward Kerker coefficients traversing a closed loop, $\sum_n K_n c^f_{n10} \approx 0$ and $\sum_n K_n c^b_{n10} \approx 0$. Though the Mie coefficients do not follow a complicated pattern, it is not immediately evident that the coefficients lead to transverse Kerker behavior, compared to the Kerker coefficients.

Finally, the right example shows highly forward directional scattering from the photonic nanojet, as evident by the open contours in the Kerker coefficients. Directionality is achieved through the interference of around 20 appreciable harmonics in each basis system. In the Kerker basis, the total forward arrow is substantially larger compared to the total backward arrow, indicating a strong preference for forward scattering. Furthermore, each coefficient has a similar magnitude. Therefore, there is no single harmonic dominating the side lobes. This is evident by the many similar sized side lobes seen in in the azimuthally integrated intensity. The electric and magnetic Mie coefficients follow a spiral pattern which indicates similar phase and magnitude behavior between the electric and magnetic-type coefficients. This pattern *almost* appears to satisfy Kerker's condition. Though, as evident by the nonzero backward Kerker coefficients, this system is not Kerker scattering. Furthermore, discrepancies between the electric and magnetic-type coefficients eventually cause the arrows of the two types to become out of synch. This makes the overall directionality harder to gauge. Finally, the Mie coefficients do nothing to illuminate the nature of side lobes.

Besides viewing coefficients in the complex plane, intuition can also be developed by studying the analytic form of directional fields in the Kerker basis based on the properties defined in section two. For example, equip with the Kerker basis, the two systems in the introduction can now be rewritten as

$$\begin{aligned} &System\ 1: \quad a\boldsymbol{Y}^f_{111} + b\boldsymbol{Y}^f_{210} + c\boldsymbol{Y}^f_{311} + a\boldsymbol{Y}^b_{111} - b\boldsymbol{Y}^b_{210} \\ &System\ 2: \quad a\boldsymbol{Y}^f_{111} + b\boldsymbol{Y}^f_{210} + c\boldsymbol{Y}^f_{311} + a\boldsymbol{Y}^b_{111} + b\boldsymbol{Y}^b_{210}, \end{aligned}$$

where $a, b, c \in \mathbb{Z}^+ \ll \infty$. Completely by inspection, the following can be concluded about the two systems: First, both systems are always forward dominant, regardless of the choice of $a, b$, or $c$. The forward-to-backward ratios are proportional to $\frac{\|a+b+c\|^2}{\|a-b\|^2}$ and $\frac{\|a+b+c\|^2}{\|a+b\|^2}$, respectively. Therefore, system 1 will always have the larger forward-to-backward ratio. Assuming $a$, $b$, and $c$ have a similar value, three lobes in the range $\theta \in [0, \pi]$ or less are expected for both systems (two lobes in the forward hemisphere and the other lobe in the backward hemisphere)[§5]. System 2 has constructive backward interference, $a + b$, which favors lobes near the exact backward direction. Alternatively, system 1 has destructive backward interference, $a - b$, which favors pushing power away from $\theta = \pi$ and into the sides. If either $a$, $b$, or $c$ are strongly dominant, then the system degenerates to more closely mimic the corresponding dominant Kerker harmonic. Side lobe predictions will change accordingly. Finally, a forward Kerker condition is clearly observed in both systems because $\boldsymbol{Y}^b_{311} = 0$, regardless of the choice of $a, b$, or $c$. Though, neither system is fully forward Kerker as $\boldsymbol{Y}^b_{111}$ and $\boldsymbol{Y}^b_{210}$ are nonzero. System 1 has the potential to be generalized Kerker if $a = b$ (suitably normalized). System 2 can only be generalized directional since $a$ and $b$ are constrained to the positive integers. We encourage the reader to return to the introduction and attempt to formulate these conclusions from the Mie framework. Examples of the two systems for different values of $a$, $b$, and $c$ are presented in the supplementary information.

**CONCLUSION**

We propose a linear transform to convert the atom-like vector spherical harmonics found in Mie theory to forward and backward directional vector harmonics and show the use of this method to understand directional scattering/emissive systems. The directional harmonics, termed the Kerker harmonics, have a simple far field expression governed primarily by the Kerker polar angle functions. These functions have a clear notion of primary and side lobes, weak coupling between forward and backward types, and coupling between same type harmonics controls directivity. The resulting azimuthally integrated and exact forward or backward intensity both have a simple analytic form which leads to intuitive definitions of Kerker, generalized Kerker, transverse Kerker, and highly directional scattering /emission as open and closed loop contours

---

[§5] This prediction is according to the side lobe formula proposed in the second section (valid for a single harmonic), the rule-of-thumb of negligible opposite type coupling, and knowledge of the general behavior of same type coupling. An inspective approach to understanding side lobes is intended to provide an educated guess based on the behavior of the Kerker harmonics. You cannot, in general, completely accurately predict side lobe behavior without performing the calculation. For example, in general, interference between forward and backward harmonics can alter side lobes. This effect is more appreciable for lobes of different type both near $\theta = \pi/2$. So, care must be taken to infer the exact location and strength of lobes near this direction. Other features, such as the existence of the Kerker effect for a particular harmonic, can be concluded completely by inspection.

of Kerker coefficients in the complex plane. Total power flow is related to the arc length of these coefficients. This provides a simple definition for the condition of theoretically maximal directivity. Examples of a Kerker, generalized Kerker, transverse Kerker, and highly directional system are shown to be more conceptually intuitive in the Kerker basis compared to the Mie basis when viewed in the complex plane. These examples explore the use of this transform in both scattering and emissive systems ranging from sub-wavelength to larger-than-wavelength size regimes (e.g., 20 appreciable harmonics).

### Additional Information

Includes a component expansion of the Kerker harmonics, derivations of all major equations, and examples of the two proposed systems for different values of $a$, $b$, and $c$.

### Author contributions statement

P.R.W. devised the research idea and developed the theoretical framework. H.A.A. oversaw the project progress. All authors contributed to writing and editing the manuscript.

### Funding Sources

This work was supported by the Army Research Office under MURI Grant W911NF-18-1-0240.

### Competing interests

The authors declare no competing financial interests.

# Supplementary Information for the Kerker Transform: Expanding Fields in a Discrete Basis of Directional Harmonics

## I. CURL OF THE KERKER HARMONICS

The Kerker basis is related to the vector spherical harmonics through the transform

$$\boldsymbol{Y}^{t}_{nmp}(\boldsymbol{r}) = (-1)^{t(n+m+1)}(i)^{n}\left(\boldsymbol{\Psi}^{\mathrm{M}}_{nmp}(\boldsymbol{r}) + (-1)^{t-p} i\boldsymbol{\Psi}^{\mathrm{E}}_{nm1-p}(\boldsymbol{r})\right) \qquad I.1$$

where $t = 0 = f$ and $t = 1 = b$. The curl of equation I.1 is then,

$$\nabla \times \boldsymbol{Y}^{t}_{nmp}(\boldsymbol{r}) = (-1)^{t(n+m+1)}(i)^{n}\left(\nabla \times \boldsymbol{\Psi}^{\mathrm{M}}_{nmp}(\boldsymbol{r}) + (-1)^{t-p} i\nabla \times \boldsymbol{\Psi}^{\mathrm{E}}_{nm1-p}(\boldsymbol{r})\right) \qquad I.2$$

Given the property of the Mie harmonics, $\nabla \times \boldsymbol{\Psi}^{t}_{nmp}(\boldsymbol{r}) = k\boldsymbol{\Psi}^{1-t}_{nmp}(\boldsymbol{r})$ where $t = 0 = M$ and $t = 1 = E$, Equation I.2 is rewritten as

$$\begin{aligned}
&\nabla \times \boldsymbol{Y}^{t}_{nmp}(\boldsymbol{r}) = (-1)^{t(n+m+1)}(i)^{n}k\left(\boldsymbol{\Psi}^{\mathrm{E}}_{nmp}(\boldsymbol{r}) + (-1)^{t-p} i\boldsymbol{\Psi}^{\mathrm{M}}_{nm1-p}(\boldsymbol{r})\right) \\
&= (-1)^{t(n+m+1)}(i)^{n}(-1)^{t-p}ki\left(\boldsymbol{\Psi}^{\mathrm{M}}_{nm1-p}(\boldsymbol{r}) - (-1)^{-(t-p)} i\boldsymbol{\Psi}^{\mathrm{E}}_{nmp}(\boldsymbol{r})\right) \\
&= (-1)^{t(n+m+1)}(i)^{n}(-1)^{t-p}ki\left(\boldsymbol{\Psi}^{\mathrm{M}}_{nm1-p}(\boldsymbol{r}) - (-1)^{t+p} i\boldsymbol{\Psi}^{\mathrm{E}}_{nmp}(\boldsymbol{r})\right) \\
&= (-1)^{t-p}ki\boldsymbol{Y}^{t}_{nm1-p}(\boldsymbol{r})
\end{aligned} \qquad I.3$$

Note that the $(-1)^{-(t-p)}$ inside the parentheses takes the same values as $(-1)^{t+p}$. Both are $-1$ when the values are opposite and 1 when the values are the same.

## II. FORM OF EM FIELDS UNDER THE KERKER EXPANSION

The electric and magnetic field expansions under the Mie framework are

$$\begin{aligned}
\boldsymbol{E}(\boldsymbol{r}) &= \sum_{n=1}^{\infty}\sum_{m=0}^{n}\sum_{p=0}^{1} c^{M}_{nmp}\boldsymbol{\Psi}^{\mathrm{M}}_{nmp} + c^{E}_{nmp}\boldsymbol{\Psi}^{\mathrm{E}}_{nmp} \\
\boldsymbol{H}(\boldsymbol{r}) &= \frac{-ik}{\mu\omega}\sum_{n=1}^{\infty}\sum_{m=0}^{n}\sum_{p=0}^{1} c^{E}_{nmp}\boldsymbol{\Psi}^{\mathrm{M}}_{nmp} + c^{M}_{nmp}\boldsymbol{\Psi}^{\mathrm{E}}_{nmp},
\end{aligned} \qquad II.2$$

where $c^{E}_{nmp}$ and $c^{M}_{nmp}$ are the electric and magnetic-type scattering coefficients, respectively.

The field expansions under the Kerker framework are

$$\begin{aligned}
\boldsymbol{E}(\boldsymbol{r}) &= \sum_{n=1}^{\infty}\sum_{m=0}^{n}\sum_{p=0}^{1} c^{f}_{nmp}\boldsymbol{Y}^{f}_{nmp}(\boldsymbol{r}) + c^{b}_{nmp}\boldsymbol{Y}^{b}_{nmp}(\boldsymbol{r}) \\
\boldsymbol{H}(\boldsymbol{r}) &= \frac{k}{\mu\omega}\sum_{n=1}^{\infty}\sum_{m=0}^{n}\sum_{p=0}^{1}(-1)^{p}\left(c^{f}_{nmp}\boldsymbol{Y}^{f}_{nm1-p}(\boldsymbol{r}) - c^{b}_{nmp}\boldsymbol{Y}^{b}_{nm1-p}(\boldsymbol{r})\right),
\end{aligned} \qquad II.2$$

where $c^{f}_{nmp}$ and $c^{b}_{nmp}$ are the forward and backward scattering coefficients, respectively. The Kerker basis is related to the vector spherical harmonics through the transform

$$\begin{aligned}
&c_{n}\boldsymbol{Y}^{t}_{nmp}(\boldsymbol{r}) = (-1)^{t(n+m+1)}(i)^{n}\left(\boldsymbol{\Psi}^{\mathrm{M}}_{nmp}(\boldsymbol{r}) + (-1)^{t-p} i\boldsymbol{\Psi}^{\mathrm{E}}_{nm1-p}(\boldsymbol{r})\right) \\
&c^{t}_{nmp} = \tfrac{1}{2}(-1)^{t(n-m-1)}(-i)^{n}\left(c^{M}_{nmp} + (-1)^{1-t-p} ic^{E}_{nm1-p}\right),
\end{aligned} \qquad II.3$$

where $t = 0 = f$ and $t = 1 = b$. Then

$$c^t_{nmp}\boldsymbol{\Upsilon}^t_{nmp}(\boldsymbol{r}) = \frac{1}{2}\left(\boldsymbol{\Psi}^M_{nmp}(\boldsymbol{r}) + (-1)^{t-p} i\boldsymbol{\Psi}^E_{nm1-p}(\boldsymbol{r})\right)\left(c^M_{nmp} + (-1)^{1-t-p} i c^E_{nm1-p}\right)$$
$$= \frac{1}{2}\begin{pmatrix} c^M_{nmp}\boldsymbol{\Psi}^M_{nmp}(\boldsymbol{r}) + c^E_{nm1-p}\boldsymbol{\Psi}^E_{nm1-p}(\boldsymbol{r}) \\ +(-1)^{1-t-p} i c^E_{nm1-p}\boldsymbol{\Psi}^M_{nmp}(\boldsymbol{r}) + (-1)^{t-p} i c^M_{nmp}\boldsymbol{\Psi}^E_{nm1-p}(\boldsymbol{r}) \end{pmatrix}, \qquad II.4$$

where the following simplifications were used: $(-1)^{t(n+m+1)}(-1)^{t(n-m-1)} = (-1)^{2tn} = 1$, $(-i)^n(i)^n = (-1)^n(i)^n = (-1)^{2n} = 1$, and $(-1)^{t-p}(-1)^{1-t-p} = (-1)^{1-2p} = -1$.

Expanding equation I.4 based on type gives,

$$c^t_{nmp}\boldsymbol{\Upsilon}^t_{nmp}(\boldsymbol{r}) + c^{1-t}_{nmp}\boldsymbol{\Upsilon}^{1-t}_{nmp}(\boldsymbol{r}) =$$
$$\frac{1}{2}\begin{pmatrix} 2c^M_{nmp}\boldsymbol{\Psi}^M_{nmp}(\boldsymbol{r}) + 2c^E_{nm1-p}\boldsymbol{\Psi}^E_{nm1-p}(\boldsymbol{r}) \\ +(-1)^{1-t-p} i c^E_{nm1-p}\boldsymbol{\Psi}^M_{nmp}(\boldsymbol{r}) + (-1)^{t-p} i c^M_{nmp}\boldsymbol{\Psi}^E_{nm1-p}(\boldsymbol{r}) \\ +(-1)^{t-p} i c^E_{nm1-p}\boldsymbol{\Psi}^M_{nmp}(\boldsymbol{r}) + (-1)^{1-t-p} i c^M_{nmp}\boldsymbol{\Psi}^E_{nm1-p}(\boldsymbol{r}) \end{pmatrix} \qquad II.5$$
$$= c^M_{nmp}\boldsymbol{\Psi}^M_{nmp}(\boldsymbol{r}) + c^E_{nm1-p}\boldsymbol{\Psi}^E_{nm1-p}(\boldsymbol{r})$$

where the simplification $(-1)^{t-p} + (-1)^{1-t-p} = (-1)^{-p}((-1)^t - (-1)^{-t}) = 0$ is used. Therefore,

$$\boldsymbol{E}(\boldsymbol{r}) = \sum\nolimits_{tnmp} c^t_{nmp}\boldsymbol{\Upsilon}^t_{nmp}(\boldsymbol{r}) = \sum\nolimits_{nmp} c^M_{nmp}\boldsymbol{\Psi}^M_{nmp}(\boldsymbol{r}) + c^E_{nm1-p}\boldsymbol{\Psi}^E_{nm1-p}(\boldsymbol{r}) \qquad II.6$$

To solve for the magnetic field, we first expand the following term

$$c^t_{nmp}\boldsymbol{\Upsilon}^t_{nm1-p}(\boldsymbol{r}) = \frac{1}{2}\left(\boldsymbol{\Psi}^M_{nm1-p}(\boldsymbol{r}) + (-1)^{t-1+p} i\boldsymbol{\Psi}^E_{nmp}(\boldsymbol{r})\right)\left(c^M_{nmp} + (-1)^{1-t-p} i c^E_{nm1-p}\right)$$
$$= \frac{1}{2}\begin{pmatrix} c^M_{nmp}\boldsymbol{\Psi}^M_{nm1-p}(\boldsymbol{r}) - c^E_{nm1-p}\boldsymbol{\Psi}^E_{nmp}(\boldsymbol{r}) \\ +(-1)^{1-t-p} i\left(c^E_{nm1-p}\boldsymbol{\Psi}^M_{nm1-p}(\boldsymbol{r}) + c^M_{nmp}\boldsymbol{\Psi}^E_{nmp}(\boldsymbol{r})\right) \end{pmatrix}, \qquad II.7$$

where the simplifications $(-1)^{t-1+p}(-1)^{1-t-p} = 1$ and $(-1)^{t-1+p} = (-1)^{1-t-p}$ are used.

Subtracting equation II.7 based on type gives,

$$c^t_{nmp}\boldsymbol{\Upsilon}^t_{nm1-p}(\boldsymbol{r}) - c^{1-t}_{nmp}\boldsymbol{\Upsilon}^{1-t}_{nm1-p}(\boldsymbol{r}) =$$
$$\frac{1}{2}\begin{pmatrix} \left(c^M_{nmp}\boldsymbol{\Psi}^M_{nm1-p}(\boldsymbol{r}) - c^E_{nm1-p}\boldsymbol{\Psi}^E_{nmp}(\boldsymbol{r})\right) - \left(c^M_{nmp}\boldsymbol{\Psi}^M_{nm1-p}(\boldsymbol{r}) - c^E_{nm1-p}\boldsymbol{\Psi}^E_{nmp}(\boldsymbol{r})\right) \\ +(-1)^{1-t-p} i\left(c^E_{nm1-p}\boldsymbol{\Psi}^M_{nm1-p}(\boldsymbol{r}) + c^M_{nmp}\boldsymbol{\Psi}^E_{nmp}(\boldsymbol{r})\right) \\ +(-1)^{1+t-p} i\left(c^E_{nm1-p}\boldsymbol{\Psi}^M_{nm1-p}(\boldsymbol{r}) + c^M_{nmp}\boldsymbol{\Psi}^E_{nmp}(\boldsymbol{r})\right) \end{pmatrix} \qquad II.8$$
$$= (-1)^{p+t}(-i)\left(c^E_{nm1-p}\boldsymbol{\Psi}^M_{nm1-p}(\boldsymbol{r}) + c^M_{nmp}\boldsymbol{\Psi}^E_{nmp}(\boldsymbol{r})\right)$$

Correspondingly,

$$\boldsymbol{H}(\boldsymbol{r}) = \frac{k}{\mu\omega}\sum\nolimits_{nmp}(-1)^p\left(c^f_{nmp}\boldsymbol{\Upsilon}^f_{nm1-p}(\boldsymbol{r}) - c^b_{nmp}\boldsymbol{\Upsilon}^b_{nm1-p}(\boldsymbol{r})\right)$$
$$= \frac{k}{\mu\omega}\sum\nolimits_{nmp}(-1)^p(-1)^{p+0}(-i)\left(c^E_{nm1-p}\boldsymbol{\Psi}^M_{nm1-p}(\boldsymbol{r}) + c^M_{nmp}\boldsymbol{\Psi}^E_{nmp}(\boldsymbol{r})\right) \qquad II.9$$
$$= \frac{-ik}{\mu\omega}\sum\nolimits_{nmp} c^E_{nm1-p}\boldsymbol{\Psi}^M_{nm1-p}(\boldsymbol{r}) + c^M_{nmp}\boldsymbol{\Psi}^E_{nmp}(\boldsymbol{r}).$$

Note that $c^f_{nmp}\boldsymbol{\Upsilon}^f_{nm1-p}(\boldsymbol{r}) - c^b_{nmp}\boldsymbol{\Upsilon}^b_{nm1-p}(\boldsymbol{r})$ is the special case of $c^t_{nmp}\boldsymbol{\Upsilon}^t_{nm1-p}(\boldsymbol{r}) - c^{1-t}_{nmp}\boldsymbol{\Upsilon}^{1-t}_{nm1-p}(\boldsymbol{r})$ when $t = 0$.

## III. VECTOR EXPANSION OF THE KERKER HARMONICS

The Mie harmonics are expanded as

$$\boldsymbol{\Psi}^{\mathrm{M}(\nu)}_{nm{0 \atop 1}}(\boldsymbol{r}) = \frac{z^{(\nu)}_n(kr)}{\sqrt{2n(n+1)}}\left({-\sin(m\phi) \atop \cos(m\phi)} m\pi^{|m|}_n(\cos\theta)\hat{\boldsymbol{e}}_\theta - {\cos(m\phi) \atop \sin(m\phi)}\tau^{|m|}_n(\cos\theta)\hat{\boldsymbol{e}}_\phi\right)$$

$$\boldsymbol{\Psi}^{\mathrm{E}(\nu)}_{nm{0 \atop 1}} = \frac{1}{\sqrt{2n(n+1)}}\begin{pmatrix} {\cos(m\phi) \atop \sin(m\phi)} n(n+1)\frac{z^{(\nu)}_n(kr)}{kr}P^{|m|}_n(\cos\theta)\hat{\boldsymbol{e}}_r \\ +\frac{\dot{z}^{(\nu)}_n(kr)}{kr}\left({\cos(m\phi) \atop \sin(m\phi)}\tau^{|m|}_n(\cos\theta)\hat{\boldsymbol{e}}_\theta + {-\sin(m\phi) \atop \cos(m\phi)} m\pi^{|m|}_n(\cos\theta)\hat{\boldsymbol{e}}_\phi\right) \end{pmatrix} \qquad III.1$$

where

$$z_n^{(\nu)}(kr) = \begin{cases} j_n(kr) & \nu = 0, r \neq \infty \\ h_n(kr) & \nu = 1, r \neq 0 \end{cases} \qquad III.2$$

and $\dot{z}_n^{(\nu)}(kr) = \partial_{kr}\left(krz_n^{(\nu)}(kr)\right) = z_n^{(3)}(kr) + kr\partial_{kr}\left(z_n^{(3)}(kr)\right)$.

The Kerker harmonics are expanded as

$$\boldsymbol{\Upsilon}^{t(\nu)}_{nm\substack{0\\1}} = (-1)^{t(n+m+1)}(i)^n\left(\boldsymbol{\Psi}^{\mathrm{M}}_{nm\substack{0\\1}}(\boldsymbol{r}) + (-1)^{t-p}i\boldsymbol{\Psi}^{\mathrm{E}}_{nm\substack{1\\0}}(\boldsymbol{r})\right)$$

$$= \hbar^t_{nmp}\begin{bmatrix} (-1)^{t-p}n(n+1)i\frac{z_n^{(\nu)}(kr)}{kr}\begin{matrix}sin(m\phi)\\cos(m\phi)\end{matrix}P_n^{|m|}(cos\,\theta)\hat{\boldsymbol{e}}_r \\ +\left((-1)^{1-p}z_n^{(\nu)}(kr)\begin{matrix}sin(m\phi)\\cos(m\phi)\end{matrix}m\pi_n^{|m|}(cos\,\theta) + (-1)^{t-p}i\frac{\dot{z}_n^{(\nu)}(kr)}{kr}\begin{matrix}sin(m\phi)\\cos(m\phi)\end{matrix}\tau_n^{|m|}(cos\,\theta)\right)\hat{\boldsymbol{e}}_\theta \\ +\left(-z_n^{(\nu)}(kr)\begin{matrix}cos(m\phi)\\sin(m\phi)\end{matrix}\tau_n^{|m|}(cos\,\theta) + (-1)^{t-p}(-1)^p i\frac{\dot{z}_n^{(\nu)}(kr)}{kr}\begin{matrix}cos(m\phi)\\sin(m\phi)\end{matrix}m\pi_n^{|m|}(cos\,\theta)\right)\hat{\boldsymbol{e}}_\phi \end{bmatrix}$$

$$= \hbar^t_{nmp}\begin{bmatrix} (-1)^t n(n+1)i\frac{z_n^{(\nu)}(kr)}{kr}\begin{matrix}sin(m\phi)\\-cos(m\phi)\end{matrix}P_n^{|m|}(cos\,\theta)\hat{\boldsymbol{e}}_r \\ \begin{matrix}sin(m\phi)\\-cos(m\phi)\end{matrix}\left(-z_n^{(\nu)}(kr)m\pi_n^{|m|}(cos\,\theta) + i\frac{\dot{z}_n^{(\nu)}(kr)}{kr}(-1)^t\tau_n^{|m|}(cos\,\theta)\right)\hat{\boldsymbol{e}}_\theta \\ \begin{matrix}cos(m\phi)\\sin(m\phi)\end{matrix}\left(-z_n^{(\nu)}(kr)\tau_n^{|m|}(cos\,\theta) + i\frac{\dot{z}_n^{(\nu)}(kr)}{kr}(-1)^t m\pi_n^{|m|}(cos\,\theta)\right)\hat{\boldsymbol{e}}_\phi \end{bmatrix} \qquad III.3$$

$$= \hbar^t_{nmp}\begin{bmatrix} n(n+1)\,sin\left(m\phi - p\frac{\pi}{2}\right)i\frac{z_n^{(\nu)}(kr)}{kr}(-1)^t P_n^{|m|}(cos\,\theta)\hat{\boldsymbol{e}}_r \\ sin\left(m\phi - p\frac{\pi}{2}\right)\left(-z_n^{(\nu)}(kr)m\pi_n^{|m|}(cos\,\theta) + i\frac{\dot{z}_n^{(\nu)}(kr)}{kr}(-1)^t\tau_n^{|m|}(cos\,\theta)\right)\hat{\boldsymbol{e}}_\theta \\ cos\left(m\phi - p\frac{\pi}{2}\right)\left(-z_n^{(\nu)}(kr)\tau_n^{|m|}(cos\,\theta) + i\frac{\dot{z}_n^{(\nu)}(kr)}{kr}(-1)^t m\pi_n^{|m|}(cos\,\theta)\right)\hat{\boldsymbol{e}}_\phi \end{bmatrix}$$

where $\hbar^t_{nmp} = \frac{(-1)^{t(n+m+1)}(i)^n}{\sqrt{2n(n+1)}}$ and the fourth equality uses the identities $sin\left(m\phi - \frac{\pi}{2}\right) = -cos(m\phi)$ and $cos\left(m\phi - \frac{\pi}{2}\right) = sin(m\phi)$.

Thought the final form of equation III.3 is compact, it is not in terms of the Kerker polar angle functions. When expanding III.3 into the far field (section V), the Kerker polar angle functions are easily derived from the asymptotic relations of the spherical Hankel functions. But, it is useful to show that a representation based on these functions is also seen even in the near field. This can be done by rewriting equation III.3 as

$$\boldsymbol{\Upsilon}^{t(\nu)}_{nmp} = (i)^n\begin{bmatrix} (-1)^{t(n+m+1)}\sqrt{\frac{n(n+1)}{2}}sin\left(m\phi - p\frac{\pi}{2}\right)i\frac{z_n^{(\nu)}(kr)}{kr}(-1)^t P_n^{|m|}(cos\,\theta)\hat{\boldsymbol{e}}_r \\ sin\left(m\phi - p\frac{\pi}{2}\right)\sum_\rho(-1)^{\rho-t}Z_{nm}^{\rho(\nu)}(kr)X_{nm}^{\rho-t}(\theta)\,\hat{\boldsymbol{e}}_\theta \\ cos\left(m\phi - p\frac{\pi}{2}\right)\sum_\rho Z_{nm}^{\rho(\nu)}(kr)X_{nm}^{\rho-t}(\theta)\,\hat{\boldsymbol{e}}_\phi \end{bmatrix} \qquad III.4$$

where $Z_{nm}^{\rho(\nu)}(kr) = \frac{(-1)^{\rho(n+m+1)}}{2}\left(-z_n^{(\nu)}(kr) + (-1)^\rho i\frac{\partial_{kr}\left(krz_n^{(\nu)}(kr)\right)}{kr}\right)$ and $\rho \in [0,1]$. Also, $X_{nm}^{\rho-t}(\theta) = \frac{(-1)^{(\rho-t)(n+m+1)}}{\sqrt{2n(n+1)}}\left(\tau_n^m(\theta) + (-1)^{(\rho-t)}m\pi_n^m(\theta)\right)$ and $t \in [0,1]$.

## IV. KERKER BASIS IN THE FAR FIELD

In the far field, the radial dependence is described by the spherical Hankel function of the first kind and has the asymptotic forms

$$\lim_{r\to\infty} z_n^{(3)}(kr) = \frac{(-i)^n e^{ikr}}{ikr}, \quad kr \gg n^2$$
$$\lim_{r\to\infty} \partial_{kr}\left(krz_n^{(3)}(kr)\right) = \lim_{r\to\infty}\left[z_n^{(3)}(kr) + kr\partial_{kr}\left(z_n^{(3)}(kr)\right)\right] = \frac{(-i)^n e^{ikr}}{ikr} + (-i)^n e^{ikr}. \quad IV.1$$

Therefore,

$$Z_{far,nm}^{\rho(\nu)}(kr) = \frac{(-1)^{\rho(n+m+1)}}{2}\frac{(-i)^n e^{ikr}}{kr}\left(i + (-1)^\rho\left(\frac{1}{kr} + i\right)\right)$$
$$= \frac{(-1)^{\rho(n+m+1)}}{2} i \frac{(-i)^n e^{ikr}}{kr}(1 + (-1)^\rho) + O\left\{\frac{1}{(kr)^2}\right\} \quad IV.2$$
$$= \begin{cases} i\frac{(-i)^n e^{ikr}}{kr} + O\left\{\frac{1}{(kr)^2}\right\} & \rho = 0 \\ 0 & \rho = 1 \end{cases}$$

Substituting the result of IV.1 into IV.3 gives,

$$\boldsymbol{\Upsilon}_{far,nmp}^{t(1)} = (-1)^{t(n+m+1)}(i)^n\left(\boldsymbol{\Psi}_{nmp}^{M(1)}(\boldsymbol{r}) + (-1)^{t-p} i \boldsymbol{\Psi}_{nm1-p}^{E(1)}(\boldsymbol{r})\right)$$
$$=$$
$$\hbar_{nmp}^t \begin{bmatrix} \frac{(-i)^n e^{ikr}}{(kr)^2} n(n+1)\sin\left(m\phi - p\frac{\pi}{2}\right)(-1)^t P_n^{|m|}(\cos\theta)\hat{\boldsymbol{e}}_r \\ \sin\left(m\phi - p\frac{\pi}{2}\right)\left(i\frac{(-i)^n e^{ikr}}{kr} m\pi_n^{|m|}(\cos\theta) + \left(\frac{(-i)^n e^{ikr}}{(kr)^2} + i\frac{(-i)^n e^{ikr}}{kr}\right)(-1)^t\tau_n^{|m|}(\cos\theta)\right)\hat{\boldsymbol{e}}_\theta \\ \cos\left(m\phi - p\frac{\pi}{2}\right)\left(i\frac{(-i)^n e^{ikr}}{kr}\tau_n^{|m|}(\cos\theta) + \left(\frac{(-i)^n e^{ikr}}{(kr)^2} + i\frac{(-i)^n e^{ikr}}{kr}\right)(-1)^t m\pi_n^{|m|}(\cos\theta)\right)\hat{\boldsymbol{e}}_\phi \end{bmatrix}$$

$$= \hbar_{nmp}^t i\frac{(-i)^n e^{ikr}}{kr}\begin{bmatrix} n(n+1)\sin\left(m\phi - p\frac{\pi}{2}\right)(-1)^t \frac{P_n^{|m|}(\cos\theta)}{ikr}\hat{\boldsymbol{e}}_r \\ \sin\left(m\phi - p\frac{\pi}{2}\right)\left(m\pi_n^{|m|}(\cos\theta) + \left(1 + \frac{1}{ikr}\right)(-1)^t\tau_n^{|m|}(\cos\theta)\right)\hat{\boldsymbol{e}}_\theta \\ \cos\left(m\phi - p\frac{\pi}{2}\right)\left(\tau_n^{|m|}(\cos\theta) + \left(1 + \frac{1}{ikr}\right)(-1)^t m\pi_n^{|m|}(\cos\theta)\right)\hat{\boldsymbol{e}}_\phi \end{bmatrix} \quad IV.3$$

$$= \frac{(-1)^{t(n+m+1)}}{\sqrt{2n(n+1)}} i\frac{e^{ikr}}{kr}\begin{bmatrix} 0\,\hat{\boldsymbol{e}}_r \\ (-1)^t\sin\left(m\phi - p\frac{\pi}{2}\right)\left(\tau_n^{|m|}(\cos\theta) + (-1)^t m\pi_n^{|m|}(\cos\theta)\right)\hat{\boldsymbol{e}}_\theta \\ \cos\left(m\phi - p\frac{\pi}{2}\right)\left(\tau_n^{|m|}(\cos\theta) + (-1)^t m\pi_n^{|m|}(\cos\theta)\right)\hat{\boldsymbol{e}}_\phi \end{bmatrix} + O\left\{\frac{1}{(kr)^2}\right\}$$

$$= i\frac{e^{ikr}}{kr}X_{nm}^t(\theta)\begin{bmatrix} 0\,\hat{\boldsymbol{e}}_r \\ \sin\left(m\phi - p\frac{\pi}{2} + t\pi\right)\hat{\boldsymbol{e}}_\theta \\ \cos\left(m\phi - p\frac{\pi}{2}\right)\hat{\boldsymbol{e}}_\phi \end{bmatrix} + O\left\{\frac{1}{(kr)^2}\right\}$$

where, $X_{nm}^t(\theta) = \frac{(-1)^{t(n+m+1)}}{\sqrt{2n(n+1)}}\left(\tau_n^m(\theta) + (-1)^t m\pi_n^m(\theta)\right)$

## V. TIME AVERAGED FAR FIELD POYNTING VECTOR IN THE KERKER BASIS

The time averaged Poynting vector is

$$\langle \boldsymbol{S} \rangle = \frac{1}{2} Re\{\boldsymbol{E}_a \times \boldsymbol{H}_b^*\}$$

$$= \frac{1}{2} Re\left\{\left(\sum_{n=1}^{\infty} \sum_{m=0}^{n} \sum_{p=0}^{1} c_{a,nmp}^{f} \boldsymbol{\Upsilon}_{nmp}^{f(\nu)} + c_{a,nmp}^{b} \boldsymbol{\Upsilon}_{nmp}^{b(\nu)}\right) \times \right.$$

$$\left. \sum_{n'=1}^{\infty} \sum_{m'=0}^{n'} \sum_{p'=0}^{1} (-1)^{p'} \left(c_{b,n'm'p'}^{*f} \boldsymbol{\Upsilon}_{n'm'1-p'}^{*f(\nu')} - c_{b,n'm'p'}^{*b} \boldsymbol{\Upsilon}_{n'm'1-p'}^{*b(\nu')}\right)\right\}$$

$$= \frac{1}{2} Re\left\{\sum_{ll'} (-1)^{p'} \frac{k}{\mu\omega} \begin{pmatrix} c_{a,l}^{f} c_{b,l'}^{*f} \boldsymbol{\Upsilon}_{nmp}^{f(\nu)} \times \boldsymbol{\Upsilon}_{n'm'1-p'}^{*f(\nu')} \\ -c_{a,l}^{f} c_{b,l'}^{*b} \boldsymbol{\Upsilon}_{nmp}^{f(\nu)} \times \boldsymbol{\Upsilon}_{n'm'1-p'}^{*b(\nu')} \\ +c_{a,l}^{b} c_{b,l'}^{*f} \boldsymbol{\Upsilon}_{nmp}^{b(\nu)} \times \boldsymbol{\Upsilon}_{n'm'1-p'}^{*f(\nu')} \\ -c_{a,l}^{b} c_{b,l'}^{*b} \boldsymbol{\Upsilon}_{nmp}^{b(\nu)} \times \boldsymbol{\Upsilon}_{n'm'1-p'}^{*b(\nu')} \end{pmatrix}\right\} \qquad V.3$$

Equation VI.1 shows that the Poynting vector is dependent on solving $\boldsymbol{\Upsilon}_{nmp}^{t(\nu)} \times \boldsymbol{\Upsilon}_{n'm'1-p'}^{*t'(\nu')}$.

$$\boldsymbol{\Upsilon}_{nmp}^{t(\nu)} \times \boldsymbol{\Upsilon}_{n'm'1-p'}^{*t'(\nu')} = \begin{bmatrix} \hat{\boldsymbol{e}}_{\boldsymbol{r}} & \hat{\boldsymbol{e}}_{\boldsymbol{\theta}} & \hat{\boldsymbol{e}}_{\boldsymbol{\phi}} \\ \Upsilon_{nmp,r}^{t(\nu)} & \Upsilon_{nmp,\theta}^{t(\nu)} & \Upsilon_{nmp,\phi}^{t(\nu)} \\ \Upsilon_{n'm'1-p',r}^{*t'(\nu')} & \Upsilon_{n'm'1-p',\theta}^{*t'(\nu')} & \Upsilon_{n'm'1-p',\phi}^{*t'(\nu')} \end{bmatrix}$$

$$= \begin{pmatrix} \left(\Upsilon_{nmp,\theta}^{t(\nu)} \Upsilon_{n'm'1-p',\phi}^{*t'(\nu')} - \Upsilon_{nmp,\phi}^{t(\nu)} \Upsilon_{n'm'1-p',\theta}^{*t'(\nu')}\right) \hat{\boldsymbol{e}}_{\boldsymbol{r}} \\ -\left(\Upsilon_{nmp,r}^{t(\nu)} \Upsilon_{n'm'1-p',\phi}^{*t'(\nu')} - \Upsilon_{nmp,\phi}^{t(\nu)} \Upsilon_{n'm'1-p',r}^{*t'(\nu')}\right) \hat{\boldsymbol{e}}_{\boldsymbol{\theta}} \\ +\left(\Upsilon_{nmp,r}^{t(\nu)} \Upsilon_{n'm'1-p',\theta}^{*t'(\nu')} - \Upsilon_{nmp,\theta}^{t(\nu)} \Upsilon_{n'm'1-p',r}^{*t'(\nu')}\right) \hat{\boldsymbol{e}}_{\boldsymbol{\phi}} \end{pmatrix} \qquad V.2$$

In the far field, equation V.2 is simplified because $\nu = \nu' = 1$ and $\boldsymbol{\Upsilon}_{nmp,r}^{t(\nu)} = \boldsymbol{\Upsilon}_{n'm'1-p',r}^{*t'(\nu')} = 0$. Equation V.2 in the far field becomes

$$\boldsymbol{\Upsilon}_{far,nmp}^{t(1)} \times \boldsymbol{\Upsilon}_{far,n'm'1-p'}^{*t'(1)} = \left(\Upsilon_{nmp,\theta}^{t(1)} \Upsilon_{n'm'1-p',\phi}^{*t'(1)} - \Upsilon_{nmp,\phi}^{t(1)} \Upsilon_{n'm'1-p',\theta}^{*t'(1)}\right) \hat{\boldsymbol{e}}_{\boldsymbol{r}}$$

$$= i \frac{e^{ikr}}{kr} X_{nm}^{t}(\theta) \left(-i \frac{e^{-ikr}}{kr} \frac{X_{n'm'}^{t'}(\theta)}{c_{n'm'}}\right) \begin{pmatrix} sin\left(m\phi - p\frac{\pi}{2} + t\pi\right) cos\left(m'\phi - (1-p')\frac{\pi}{2}\right) \\ -\ cos\left(m\phi - p\frac{\pi}{2}\right) sin\left(m'\phi - (1-p')\frac{\pi}{2} + t'\pi\right) \end{pmatrix} \hat{\boldsymbol{e}}_{\boldsymbol{r}}$$

$$= \frac{1}{(kr)^2} X_{nm}^{t}(\theta) X_{n'm'}^{t'}(\theta) \begin{pmatrix} sin\left(m\phi - p\frac{\pi}{2} + t\pi\right) cos\left(m'\phi - (1-p')\frac{\pi}{2}\right) \\ -\ cos\left(m\phi - p\frac{\pi}{2}\right) sin\left(m'\phi - (1-p')\frac{\pi}{2} + t'\pi\right) \end{pmatrix} \hat{\boldsymbol{e}}_{\boldsymbol{r}} \qquad V.3$$

$$= \frac{1}{(kr)^2} X_{nm}^{t}(\theta) X_{n'm'}^{t'}(\theta) \begin{pmatrix} (-1)^{t} sin\left(m\phi - p\frac{\pi}{2}\right) sin\left(m'\phi + p'\frac{\pi}{2}\right) \\ (-1)^{t'} cos\left(m\phi - p\frac{\pi}{2}\right) cos\left(m'\phi + p'\frac{\pi}{2}\right) \end{pmatrix} \hat{\boldsymbol{e}}_{\boldsymbol{r}}$$

$$= \frac{1}{(kr)^2} X_{nm}^{t}(\theta) X_{n'm'}^{t'}(\theta) \begin{pmatrix} (-1)^{t}(-1)^{p'} sin\left(m\phi - p\frac{\pi}{2}\right) sin\left(m'\phi - p'\frac{\pi}{2}\right) \\ (-1)^{t'}(-1)^{p'} cos\left(m\phi - p\frac{\pi}{2}\right) cos\left(m'\phi - p'\frac{\pi}{2}\right) \end{pmatrix} \hat{\boldsymbol{e}}_{\boldsymbol{r}}$$

Substituting equation V.3 into V.1 gives,

$$\langle \boldsymbol{S} \rangle = \frac{1}{2} Re\{\boldsymbol{E}_a \times \boldsymbol{H}_b^*\} \qquad V.4$$

$$
= \frac{1}{2} Re \left\{ \frac{1}{\mu\omega k r^2} \sum_{ll'} \left( \begin{array}{l}
c^f_{a,l} c^{*f}_{b,l'} X^f_{nm}(\theta) X^f_{n'm'}(\theta) \begin{pmatrix} sin\left(m\phi - p\frac{\pi}{2}\right) sin\left(m'\phi - p'\frac{\pi}{2}\right) \\ + cos\left(m\phi - p\frac{\pi}{2}\right) cos\left(m'\phi - p'\frac{\pi}{2}\right) \end{pmatrix} \\
- c^f_{a,l} c^{*b}_{b,l'} X^f_{nm}(\theta) X^b_{n'm'}(\theta) \begin{pmatrix} sin\left(m\phi - p\frac{\pi}{2}\right) sin\left(m'\phi - p'\frac{\pi}{2}\right) \\ - cos\left(m\phi - p\frac{\pi}{2}\right) cos\left(m'\phi - p'\frac{\pi}{2}\right) \end{pmatrix} \\
+ c^b_{a,l} c^{*f}_{b,l'} X^b_{nm}(\theta) X^f_{n'm'}(\theta) \begin{pmatrix} - sin\left(m\phi - p\frac{\pi}{2}\right) sin\left(m'\phi - p'\frac{\pi}{2}\right) \\ + cos\left(m\phi - p\frac{\pi}{2}\right) cos\left(m'\phi - p'\frac{\pi}{2}\right) \end{pmatrix} \\
- c^b_{a,l} c^{*b}_{b,l'} X^b_{nm}(\theta) X^b_{n'm'}(\theta) \begin{pmatrix} - sin\left(m\phi - p\frac{\pi}{2}\right) sin\left(m'\phi - p'\frac{\pi}{2}\right) \\ - cos\left(m\phi - p\frac{\pi}{2}\right) cos\left(m'\phi - p'\frac{\pi}{2}\right) \end{pmatrix}
\end{array} \right) \right\} \hat{\boldsymbol{e}}_{\boldsymbol{r}}
$$

$$
= \frac{1}{2} Re \left\{ \frac{1}{\mu\omega k r^2} \sum_{ll'} \left( \begin{array}{l}
c^f_{a,l} c^{*f}_{b,l'} X^f_{nm}(\theta) X^f_{n'm'}(\theta) \begin{pmatrix} sin\left(m\phi - p\frac{\pi}{2}\right) sin\left(m'\phi - p'\frac{\pi}{2}\right) \\ + cos\left(m\phi - p\frac{\pi}{2}\right) cos\left(m'\phi - p'\frac{\pi}{2}\right) \end{pmatrix} \\
+ c^f_{a,l} c^{*b}_{b,l'} X^f_{nm}(\theta) X^b_{n'm'}(\theta) \begin{pmatrix} - sin\left(m\phi - p\frac{\pi}{2}\right) sin\left(m'\phi - p'\frac{\pi}{2}\right) \\ + cos\left(m\phi - p\frac{\pi}{2}\right) cos\left(m'\phi - p'\frac{\pi}{2}\right) \end{pmatrix} \\
+ c^b_{a,l} c^{*f}_{b,l'} X^b_{nm}(\theta) X^f_{n'm'}(\theta) \begin{pmatrix} - sin\left(m\phi - p\frac{\pi}{2}\right) sin\left(m'\phi - p'\frac{\pi}{2}\right) \\ + cos\left(m\phi - p\frac{\pi}{2}\right) cos\left(m'\phi - p'\frac{\pi}{2}\right) \end{pmatrix} \\
c^b_{a,l} c^{*b}_{b,l'} X^b_{nm}(\theta) X^b_{n'm'}(\theta) \begin{pmatrix} sin\left(m\phi - p\frac{\pi}{2}\right) sin\left(m'\phi - p'\frac{\pi}{2}\right) \\ + cos\left(m\phi - p\frac{\pi}{2}\right) cos\left(m'\phi - p'\frac{\pi}{2}\right) \end{pmatrix}
\end{array} \right) \right\} \hat{\boldsymbol{e}}_{\boldsymbol{r}}
$$

$$
= \frac{1}{2} Re \left\{ \frac{1}{\mu\omega k r^2} \sum_{ll'} \left( sin\left(m\phi - p\frac{\pi}{2}\right) sin\left(m'\phi - p'\frac{\pi}{2}\right) \left(c^f_{a,l} X^f_{nm}(\theta) - c^b_{a,l} X^b_{nm}(\theta)\right) \left(c^{*f}_{b,l'} X^f_{n'm'}(\theta) - c^{*b}_{b,l'} X^b_{n'm'}(\theta)\right) + \right.\right.
$$

$$
\left.\left. cos\left(m\phi - p\frac{\pi}{2}\right) cos\left(m'\phi - p'\frac{\pi}{2}\right) \left(c^f_{a,l} X^f_{nm}(\theta) + c^b_{a,l} X^b_{nm}(\theta)\right) \left(c^{*f}_{b,l'} X^f_{n'm'}(\theta) + c^{*b}_{b,l'} X^b_{n'm'}(\theta)\right) \right) \right\} \hat{\boldsymbol{e}}_{\boldsymbol{r}}
$$

Where the last equality comes from $sin\left(m\phi - p\frac{\pi}{2}\right) sin\left(m'\phi - p'\frac{\pi}{2}\right) \left(c^f_{a,l} c^{*f}_{b,l'} X^f_{nm}(\theta) X^f_{n'm'}(\theta) - c^f_{a,l} c^{*b}_{b,l'} X^f_{nm}(\theta) X^b_{n'm'}(\theta) - c^b_{a,l} c^{*f}_{b,l'} X^b_{nm}(\theta) X^f_{n'm'}(\theta) + c^b_{a,l} c^{*b}_{b,l'} X^b_{nm}(\theta) X^b_{n'm'}(\theta)\right) = sin\left(m\phi - p\frac{\pi}{2}\right) sin\left(m'\phi - p'\frac{\pi}{2}\right) \left(c^f_{a,l} X^f_{nm}(\theta) - c^b_{a,l} X^b_{nm}(\theta)\right) \left(c^{*f}_{b,l} X^f_{nm}(\theta) - c^{*b}_{b,l} X^b_{nm}(\theta)\right)$ and $cos\left(m\phi - p\frac{\pi}{2}\right) cos\left(m'\phi - p'\frac{\pi}{2}\right) \left(c^f_{a,l} c^{*f}_{b,l'} X^f_{nm}(\theta) X^f_{n'm'}(\theta) + c^f_{a,l} c^{*b}_{b,l'} X^f_{nm}(\theta) X^b_{n'm'}(\theta) + c^b_{a,l} c^{*f}_{b,l'} X^b_{nm}(\theta) X^f_{n'm'}(\theta) + c^b_{a,l} c^{*b}_{b,l'} X^b_{nm}(\theta) X^b_{n'm'}(\theta)\right) = cos\left(m\phi - p\frac{\pi}{2}\right) cos\left(m'\phi - p'\frac{\pi}{2}\right) \left(c^f_{a,l} X^f_{nm}(\theta) + c^b_{a,l} X^b_{nm}(\theta)\right) \left(c^{*f}_{b,l} X^f_{n'm'}(\theta) + c^{*b}_{b,l} X^b_{n'm'}(\theta)\right)$

If $\boldsymbol{E}_a = \boldsymbol{E}_b = \boldsymbol{E}$, then equation V.4 can be further simplified to

$$
\langle \boldsymbol{S}_{far} \rangle = \frac{1}{2} \frac{1}{\mu\omega k r^2} \left( \|A(\phi, \theta)\|^2 + \|B(\phi, \theta)\|^2 \right) \hat{\boldsymbol{e}}_r \qquad V.5
$$

where

$$
\begin{aligned}
A(\theta, \phi) &= \sum_{nmp} sin\left(m\phi - p\frac{\pi}{2}\right) \left(c^f_{nmp} X^f_{nm}(\theta) - c^b_{nmp} X^b_{nm}(\theta)\right) \\
B(\theta, \phi) &= \sum_{nmp} cos\left(m\phi - p\frac{\pi}{2}\right) \left(c^f_{nmp} X^f_{nm}(\theta) + c^b_{nmp} X^b_{nm}(\theta)\right) \qquad V.6
\end{aligned}
$$

## VI. FAR FIELD AZIMUTHALLY INTEGRATED INTENSITY IN THE KERKER BASIS

Starting from the far field Poynting vector,

$$\langle \boldsymbol{S}_{far}\rangle = \frac{1}{2}Re\{\boldsymbol{E}_a \times \boldsymbol{H}_a^*\} =$$

$$\frac{1}{2}Re\left\{\frac{1}{\mu\omega kr^2}\sum_{ll'}\left(\begin{array}{l} c_l^f c_{l'}^{*f} X_{nm}^f(\theta) X_{n'm'}^f(\theta)\begin{pmatrix} \sin\left(m\phi - p\frac{\pi}{2}\right)\sin\left(m'\phi - p'\frac{\pi}{2}\right) \\ +\cos\left(m\phi - p\frac{\pi}{2}\right)\cos\left(m'\phi - p'\frac{\pi}{2}\right)\end{pmatrix} \\ +c_l^f c_{l'}^{*b} X_{nm}^f(\theta) X_{n'm'}^b(\theta)\begin{pmatrix} -\sin\left(m\phi - p\frac{\pi}{2}\right)\sin\left(m'\phi - p'\frac{\pi}{2}\right) \\ +\cos\left(m\phi - p\frac{\pi}{2}\right)\cos\left(m'\phi - p'\frac{\pi}{2}\right)\end{pmatrix} \\ +c_l^b c_{l'}^{*f} X_{nm}^b(\theta) X_{n'm'}^f(\theta)\begin{pmatrix} -\sin\left(m\phi - p\frac{\pi}{2}\right)\sin\left(m'\phi - p'\frac{\pi}{2}\right) \\ +\cos\left(m\phi - p\frac{\pi}{2}\right)\cos\left(m'\phi - p'\frac{\pi}{2}\right)\end{pmatrix} \\ c_l^b c_{l'}^{*b} X_{nm}^b(\theta) X_{n'm'}^b(\theta)\begin{pmatrix} \sin\left(m\phi - p\frac{\pi}{2}\right)\sin\left(m'\phi - p'\frac{\pi}{2}\right) \\ +\cos\left(m\phi - p\frac{\pi}{2}\right)\cos\left(m'\phi - p'\frac{\pi}{2}\right)\end{pmatrix} \end{array}\right)\right\}\hat{\boldsymbol{e}}_{\boldsymbol{r}} \qquad \textit{VI.1}$$

The azimuthal integral of equation VI.1 makes use of the following orthogonality relations,

$$\begin{aligned} &\int_0^{2\pi}\cos(m\phi)\sin(m'\phi)\,\partial\phi = 0 \\ &\int_0^{2\pi}\cos(m\phi)\cos(m'\phi)\,\partial\phi = \begin{cases} 2\pi & m = m' = 0 \\ \pi & m = m' \neq 0 \\ 0 & m \neq m' \end{cases} \\ &\int_0^{2\pi}\sin(m\phi)\sin(m'\phi)\,\partial\phi = \begin{cases} 0 & m = m' = 0 \\ \pi & m = m' \neq 0 \\ 0 & m \neq m'. \end{cases} \end{aligned} \qquad \textit{VI.2}$$

From equation VI.2 we see all nonzero integral require $m = m'$. Under this condition, starting with $m = 0$, the azimuthal integral of equation VI.1 becomes

$$\int_0^{2\pi} r^2\partial\phi\langle\boldsymbol{S}_{far}\rangle\delta_{m0}\cdot\hat{\boldsymbol{e}}_{\boldsymbol{r}}$$

$$= \int_0^{2\pi} r^2\partial\phi\,\frac{1}{2}Re\left\{\frac{1}{\mu\omega kr^2}\sum_{ll'}\left(\begin{array}{l} c_{n0p}^f c_{n'0p'}^{*f} X_{n0}^f(\theta) X_{n'0}^f(\theta)\begin{pmatrix} \sin\left(-p\frac{\pi}{2}\right)\sin\left(-p'\frac{\pi}{2}\right) \\ +\cos\left(-p\frac{\pi}{2}\right)\cos\left(-p'\frac{\pi}{2}\right)\end{pmatrix} \\ +c_{n0p}^f c_{n'0p'}^{*b} X_{n0}^f(\theta) X_{n'0}^b(\theta)\begin{pmatrix} -\sin\left(-p\frac{\pi}{2}\right)\sin\left(-p'\frac{\pi}{2}\right) \\ +\cos\left(-p\frac{\pi}{2}\right)\cos\left(-p'\frac{\pi}{2}\right)\end{pmatrix} \\ +c_{n0p}^b c_{n'0p'}^{*f} X_{n0}^b(\theta) X_{n'0}^f(\theta)\begin{pmatrix} -\sin\left(-p\frac{\pi}{2}\right)\sin\left(-p'\frac{\pi}{2}\right) \\ +\cos\left(-p\frac{\pi}{2}\right)\cos\left(-p'\frac{\pi}{2}\right)\end{pmatrix} \\ c_{n0p}^b c_{n'0p'}^{*b} X_{n0}^b(\theta) X_{n'0}^b(\theta)\begin{pmatrix} \sin\left(-p\frac{\pi}{2}\right)\sin\left(-p'\frac{\pi}{2}\right) \\ +\cos\left(-p\frac{\pi}{2}\right)\cos\left(-p'\frac{\pi}{2}\right)\end{pmatrix} \end{array}\right)\right\} \qquad \textit{VI.3}$$

$$= \pi r^2 Re\left\{\frac{1}{\mu\omega k r^2}\sum_{ll'}\left(\begin{array}{l} c^f_{n0p}c^{*f}_{n'0p'}X^f_{n0}(\theta)X^f_{n'0}(\theta)\begin{pmatrix} sin\left(-p\frac{\pi}{2}\right)sin\left(-p'\frac{\pi}{2}\right) \\ +cos\left(-p\frac{\pi}{2}\right)cos\left(-p'\frac{\pi}{2}\right)\end{pmatrix} \\ +c^f_{n0p}c^{*b}_{n'0p'}X^f_{n0}(\theta)X^b_{n'0}(\theta)\begin{pmatrix} -sin\left(-p\frac{\pi}{2}\right)sin\left(-p'\frac{\pi}{2}\right) \\ +cos\left(-p\frac{\pi}{2}\right)cos\left(-p'\frac{\pi}{2}\right)\end{pmatrix} \\ +c^b_{n0p}c^{*f}_{n'0p'}X^b_{n0}(\theta)X^f_{n'0}(\theta)\begin{pmatrix} -sin\left(-p\frac{\pi}{2}\right)sin\left(-p'\frac{\pi}{2}\right) \\ +cos\left(-p\frac{\pi}{2}\right)cos\left(-p'\frac{\pi}{2}\right)\end{pmatrix} \\ c^b_{n0p}c^{*b}_{n'0p'}X^b_{n0}(\theta)X^b_{n'0}(\theta)\begin{pmatrix} sin\left(-p\frac{\pi}{2}\right)sin\left(-p'\frac{\pi}{2}\right) \\ +cos\left(-p\frac{\pi}{2}\right)cos\left(-p'\frac{\pi}{2}\right)\end{pmatrix}\end{array}\right)\right\}$$

Evaluating the trigonometric functions in equation VI.3 gives

| | $sin\left(-p\frac{\pi}{2}\right)sin\left(-p'\frac{\pi}{2}\right)$ $+cos\left(-p\frac{\pi}{2}\right)cos\left(-p'\frac{\pi}{2}\right)$ | $-sin\left(-p\frac{\pi}{2}\right)sin\left(-p'\frac{\pi}{2}\right)$ $+cos\left(-p\frac{\pi}{2}\right)cos\left(-p'\frac{\pi}{2}\right)$ |
|---|---|---|
| $p=0, p'=0$ | 1 | 1 |
| $p=0, p'=1$ | 0 | 0 |
| $p=1, p'=0$ | 0 | 0 |
| $p=1, p'=1$ | 1 | −1 |

*VI.4*

Therefore, $p = p'$ are the only nonzero terms and equation VI.3 can be written as

$$\int_0^{2\pi} r^2\partial\phi\langle \boldsymbol{S}_{far}\rangle\delta_{m0}\cdot\hat{\boldsymbol{e}}_{\boldsymbol{r}}$$

$$= \pi r^2 Re\left\{\frac{1}{\mu\omega k r^2}\sum_{ll'}\begin{pmatrix} c^f_{n00}c^{*f}_{n'00}X^f_{n0}(\theta)X^f_{n'0}(\theta)+c^f_{n01}c^{*f}_{n'01}X^f_{n0}(\theta)X^f_{n'0}(\theta) \\ +c^f_{n00}c^{*b}_{n'00}X^f_{n0}(\theta)X^b_{n'0}(\theta)-c^f_{n01}c^{*b}_{n'01}X^f_{n0}(\theta)X^b_{n'0}(\theta) \\ +c^b_{n00}c^{*f}_{n'00}X^b_{n0}(\theta)X^f_{n'0}(\theta)-c^b_{n01}c^{*f}_{n'01}X^b_{n0}(\theta)X^f_{n'0}(\theta) \\ +c^b_{n00}c^{*b}_{n'00}X^b_{n0}(\theta)X^b_{n'0}(\theta)+c^b_{n01}c^{*b}_{n'01}X^b_{n0}(\theta)X^b_{n'0}(\theta)\end{pmatrix}\right\} \qquad VI.5$$

Note that $X^t_{n0}(\theta) = \frac{(-1)^{t(n+1)}}{\sqrt{2n(n+1)}}\tau^0_n(\theta)$ and $c^t_{n0p} = \frac{1}{2}(-1)^{t(n-1)}(-i)^n\left(c^M_{n0p} + (-1)^{1-t-p}ic^E_{n01-p}\right)$. Therefore, $c^t_{n0p}X^t_{n0}(\theta) = \frac{(-i)^n}{2\sqrt{2n(n+1)}}\left(c^M_{n0p} + (-1)^{1-t-p}ic^E_{n01-p}\right)$. Also $c^M_{n01} = c^E_{n01} = 0$, so $c^t_{n00}X^t_{n0}(\theta) = \frac{(-i)^n}{2\sqrt{2n(n+1)}}(c^M_{n00})$ and $c^t_{n01}X^t_{n0}(\theta) = \frac{(-i)^n}{2\sqrt{2n(n+1)}}((-1)^t ic^E_{n00})$. Given that $c^t_{n00}X^t_{n0}(\theta)$ is not t-dependent, then $c^t_{n00}X^t_{n0}(\theta)c^{*t'}_{n00}X^{t'}_{n0}(\theta) = \frac{c^M_{n00}c^{*M}_{n'00}}{4\sqrt{2n(n+1)}\sqrt{2n'(n'+1)}} = A_{nn'}$. Also, $c^t_{n01}X^t_{n0}(\theta)c^{*t'}_{n01}X^{t'}_{n0}(\theta) = \frac{(-1)^{t+t'}c^E_{n00}c^{*E}_{n'00}}{4\sqrt{2n(n+1)}\sqrt{2n'(n'+1)}}$. Let $B_{nn'} = \frac{c^E_{n00}c^{*E}_{n'00}}{4\sqrt{2n(n+1)}\sqrt{2n'(n'+1)}} = c^t_{n01}X^t_{n0}(\theta)c^{*t}_{n01}X^t_{n0}(\theta)$. Then equation VI.5 can be written as

$$\int_0^{2\pi} r^2\partial\phi\langle \boldsymbol{S}_{far}\rangle\delta_{m0}\cdot\hat{\boldsymbol{e}}_{\boldsymbol{r}}$$

$$= \pi r^2 Re\left\{\frac{1}{\mu\omega k r^2}\sum_{nn'}4A_{nn'} + 4B_{nn'}\right\}$$

$$= \pi r^2 Re\left\{\frac{1}{\mu\omega k r^2}\sum_{nn't}2c^t_{n00}X^t_{n0}(\theta)c^{*t}_{n00}X^t_{n0}(\theta) + 2c^t_{n01}X^t_{n0}(\theta)c^{*t}_{n01}X^t_{n0}(\theta)\right\} \qquad VI.5$$

$$= \pi r^2 Re\left\{\frac{1}{\mu\omega k r^2}\sum_{nn'tp}c^t_{n0p}X^t_{n0}(\theta)c^{*t}_{n0p}X^t_{n0}(\theta)\right\}$$

$$= 2\pi r^2 Re\left\{\frac{1}{\mu\omega k r^2}\sum_{tp}\left\|\sum_n c^t_{n00}X^t_{n0}(\theta)\right\|^2\right\}$$

When $m \neq 0$, the azimuthal integral of equation VI.1 becomes

$$\int_0^{2\pi} r^2 \partial\phi \langle \boldsymbol{S}_{far}\rangle \cdot \hat{\boldsymbol{e}}_{\boldsymbol{r}} = \int_0^{2\pi} r^2 \partial\phi \frac{1}{2}$$

$$Re\left\{\frac{1}{\mu\omega k r^2}\sum_{ll'}\delta_{mm}\begin{pmatrix} c_l^f c_{l'}^{*f} X_{nm}^f(\theta) X_{n'm}^f(\theta)\begin{pmatrix} \sin\left(m\phi - p\frac{\pi}{2}\right)\sin\left(m\phi - p'\frac{\pi}{2}\right) \\ +\cos\left(m\phi - p\frac{\pi}{2}\right)\cos\left(m\phi - p'\frac{\pi}{2}\right)\end{pmatrix} \\ +c_l^f c_{l'}^{*b} X_{nm}^f(\theta) X_{n'm}^b(\theta)\begin{pmatrix} -\sin\left(m\phi - p\frac{\pi}{2}\right)\sin\left(m\phi - p'\frac{\pi}{2}\right) \\ +\cos\left(m\phi - p\frac{\pi}{2}\right)\cos\left(m\phi - p'\frac{\pi}{2}\right)\end{pmatrix} \\ +c_l^b c_{l'}^{*f} X_{nm}^b(\theta) X_{n'm}^f(\theta)\begin{pmatrix} -\sin\left(m\phi - p\frac{\pi}{2}\right)\sin\left(m\phi - p'\frac{\pi}{2}\right) \\ +\cos\left(m\phi - p\frac{\pi}{2}\right)\cos\left(m\phi - p'\frac{\pi}{2}\right)\end{pmatrix} \\ c_l^b c_{l'}^{*b} X_{nm}^b(\theta) X_{n'm}^b(\theta)\begin{pmatrix} \sin\left(m\phi - p\frac{\pi}{2}\right)\sin\left(m\phi - p'\frac{\pi}{2}\right) \\ +\cos\left(m\phi - p\frac{\pi}{2}\right)\cos\left(m\phi - p'\frac{\pi}{2}\right)\end{pmatrix}\end{pmatrix}\right\} \quad VI.6$$

Then, by the orthogonality relations in VI.2,

$$\begin{aligned}&\int_0^{2\pi} r^2 \partial\phi \langle \boldsymbol{S}_{far}\rangle \cdot \hat{\boldsymbol{e}}_{\boldsymbol{r}} \\ &= \frac{1}{2} Re\left\{\frac{1}{\mu\omega k}\sum_{ll'} 2\pi \delta_{mm'}\delta_{pp'}\left(c_{nmp}^f c_{n'mp}^{*f} X_{nm}^f(\theta) X_{n'm}^f(\theta) + c_{nmp}^b c_{n'mp}^{*b} X_{nm}^b(\theta) X_{n'm}^b(\theta)\right)\right\} \\ &= \frac{\pi}{\mu\omega k}\sum_{tmp}\left\|\sum_n c_{nmp}^t X_{nm}^t(\theta)\right\|^2\end{aligned} \quad VI.7$$

## VII. ORTHOGONALITY OF THE POLAR ANGLE FUNCTION

The Kerker polar angle functions are

$$X_{nm}^{\rho-t}(\theta) = \frac{(-1)^{(\rho-t)(n+m+1)}}{\sqrt{2n(n+1)}}\left(\tau_n^m(\theta) + (-1)^{(\rho-t)} m\pi_n^m(\theta)\right) \quad VII.1$$

where

$$\tau_n^m(\theta) = \frac{d}{d\theta} P_n^m(\cos(\theta))$$
$$\pi_n^m(\theta) = \frac{P_n^m(\cos(\theta))}{\sin(\theta)} \quad . \quad VII.2$$

Then,

$$\begin{aligned}&\int_0^{\pi} \partial\theta \sin(\theta)\, X_{nm}^{\rho-t}(\theta) X_{n'm}^{\rho'-t'}(\theta) \\ &= \int_0^{\pi} \partial\theta \sin(\theta) \frac{(-1)^{(\rho-t)(n+m+1)}}{\sqrt{2n(n+1)}} \frac{(-1)^{(\rho'+t')(n'+m+1)}}{\sqrt{2n'(n'+1)}}\begin{pmatrix}\left(\tau_n^m(\theta) + (-1)^{(\rho-t)} m\pi_n^m(\theta)\right) \\ \times\left(\tau_{n'}^m(\theta) + (-1)^{(\rho'-t')} m\pi_{n'}^m(\theta)\right)\end{pmatrix} \\ &= \frac{(-1)^{(\rho-t)(n+m+1)}}{\sqrt{2n(n+1)}} \frac{(-1)^{(\rho'-t')(n'+m+1)}}{\sqrt{2n'(n'+1)}} \int_0^{\pi} \partial\theta \sin(\theta)\left(\mathcal{A}_{nn'm}^{\rho\rho'}(\theta) + \mathcal{B}_{nn'm}^{\rho\rho'}(\theta)\right)\end{aligned} \quad VII.3$$

where

$$\mathcal{A}_{nn'm}^{\rho\rho'}(\theta) = \tau_n^m(\theta)\tau_{n'}^m(\theta) + (-1)^{(\rho+\rho'-t-t')} m^2 \pi_n^m(\theta)\pi_{n'}^m(\theta)$$
$$\mathcal{B}_{nn'm}^{\rho\rho'}(\theta) = (-1)^{(\rho'-t')}\tau_n^m(\theta) m\pi_{n'}^m(\theta) + (-1)^{(\rho-t)} m\pi_n^m(\theta)\tau_{n'}^m(\theta) \quad VII.4$$

Equation VI.4 comes from

$$\begin{aligned}&\left(\tau_n^m(\theta) + (-1)^{(\rho-t)} m\pi_n^m(\theta)\right)\left(\tau_{n'}^m(\theta) + (-1)^{(\rho'-t')} m\pi_{n'}^m(\theta)\right) \\ &= \begin{pmatrix} \tau_n^m(\theta)\tau_{n'}^m(\theta) + (-1)^{(\rho'-t')}\tau_n^m(\theta) m\pi_{n'}^m(\theta) \\ +(-1)^{(\rho-t)} m\pi_n^m(\theta)\tau_{n'}^m(\theta) + (-1)^{(\rho+\rho'-t-t')} m\pi_n^m(\theta) m\pi_{n'}^m(\theta)\end{pmatrix}\end{aligned} \quad VII.5$$

The solution to equation VII.3 is dependent on the solutions to the integral of the equations in VII.4.

The first integral is

$$\int_0^\pi \partial\theta \sin(\theta)\, \mathcal{A}_{nn'm}^{\rho\rho'}(\theta) = \int_0^\pi \partial\theta \sin(\theta) \left(\tau_n^m(\theta)\tau_{n'}^m(\theta) + (-1)^{(\rho+\rho'-t-t')} m^2 \pi_n^m(\theta)\pi_{n'}^m(\theta)\right) =$$
$$\int_0^\pi \partial\theta \sin(\theta) \left(\frac{\partial}{\partial\theta} P_n^m(\cos(\theta)) \frac{\partial}{\partial\theta} P_{n'}^m(\cos(\theta)) + (-1)^{(\rho+\rho'-t-t')} m^2 \frac{P_n^m(\cos(\theta))}{\sin(\theta)} \frac{P_{n'}^m(\cos(\theta))}{\sin(\theta)}\right) \quad VII.6$$

Letting

$$u = \sin(\theta) \frac{\partial}{\partial\theta} P_n^m(\cos(\theta)) \qquad v = P_{n'}^m(\cos(\theta))$$
$$\frac{\partial u}{\partial\theta} = \frac{\partial}{\partial\theta}\left(\sin(\theta) \frac{\partial}{\partial\theta} P_n^m(\cos(\theta))\right) \quad \frac{\partial v}{\partial\theta} = \frac{\partial}{\partial\theta} P_{n'}^m(\cos(\theta)) \quad VII.7$$

Then, integrating the first term in VII.6 by parts gives

$$\int_0^\pi \partial\theta \sin(\theta) \frac{\partial}{\partial\theta} P_n^m(\cos(\theta)) \frac{\partial}{\partial\theta} P_{n'}^m(\cos(\theta))$$
$$= \sin(\theta) \frac{\partial P_n^m(\cos(\theta))}{\partial\theta} P_{n'}^m(\cos(\theta))\Big|_0^\pi - \int_0^\pi \partial\theta P_{n'}^m(\cos(\theta)) \frac{\partial}{\partial\theta}\left(\sin(\theta) \frac{\partial}{\partial\theta} P_n^m(\cos(\theta))\right) \quad VII.8$$

The first term in VII.8 is

$$\sin(\pi) \frac{\partial P_n^m(-1)}{\partial\theta} P_{n'}^m(-1) - \sin(0) \frac{\partial P_n^m(1)}{\partial\theta} P_{n'}^m(1) = 0\ . \quad VII.9$$

Using the identity

$$\frac{\partial}{\partial\theta}\left(\sin(\theta) \frac{\partial}{\partial\theta} P_n^m(\cos(\theta))\right) = \frac{m^2 P_n^m(\cos(\theta))}{\sin(\theta)} - n(n+1)\sin(\theta)\, P_n^m(\cos(\theta)), \quad VII.10$$

the second term VII.8 is,

$$-\int_0^\pi \partial\theta P_{n'}^m(\cos(\theta)) \frac{\partial}{\partial\theta}\left(\sin(\theta) \frac{\partial}{\partial\theta} P_n^m(\cos(\theta))\right)$$
$$= -\int_0^\pi \sin(\theta)\, \partial\theta \left(\frac{m^2 P_n^m(\cos(\theta)) P_{n'}^m(\cos(\theta))}{\sin^2(\theta)} - n(n+1) P_n^m(\cos(\theta)) P_{n'}^m(\cos(\theta))\right) \quad VII.11$$

Therefore, equation VII.6 can be written as,

$$\int_0^\pi \partial\theta \sin(\theta)\, \mathcal{A}_{nn'm}^{\rho\rho'}(\theta)$$
$$= \int_0^\pi \partial\theta \sin(\theta) \begin{pmatrix} n(n+1) P_n^m(\cos(\theta)) P_{n'}^m(\cos(\theta)) \\ +\left((-1)^{(\rho+\rho'-t-t')} - 1\right) m^2 \frac{P_n^m(\cos(\theta))}{\sin(\theta)} \frac{P_{n'}^m(\cos(\theta))}{\sin(\theta)} \end{pmatrix} \quad VII.12$$
$$= n(n+1)\delta_{nn'} + \int_0^\pi \partial\theta \sin(\theta) \left((-1)^{(\rho+\rho'-t-t')} - 1\right) m^2 \frac{P_n^m(\cos(\theta))}{\sin(\theta)} \frac{P_{n'}^m(\cos(\theta))}{\sin(\theta)}$$

Likewise,

$$\int_0^\pi \partial\theta \sin(\theta)\, \mathcal{B}_{nn'm}^{\rho\rho'}(\theta)$$
$$= \int_0^\pi \partial\theta \sin(\theta) \left((-1)^{(\rho'-t')} \tau_n^m(\theta) m \pi_{n'}^m(\theta) + (-1)^{(\rho-t)} m \pi_n^m(\theta) \tau_{n'}^m(\theta)\right)$$
$$= \int_0^\pi \partial\theta \sin(\theta) \left((-1)^{(\rho'-t')} m \frac{\partial P_n^m(\cos(\theta))}{\partial\theta} \frac{P_{n'}^m(\cos(\theta))}{\sin(\theta)} + (-1)^{(\rho-t)} m \frac{\partial P_{n'}^m(\cos(\theta))}{\partial\theta} \frac{P_n^m(\cos(\theta))}{\sin(\theta)}\right) \quad VII.13$$
$$= m \int_0^\pi \partial\theta \left((-1)^{(\rho'-t')} \frac{\partial P_n^m(\cos(\theta))}{\partial\theta} P_{n'}^m(\cos(\theta)) + (-1)^{(\rho-t)} \frac{\partial P_{n'}^m(\cos(\theta))}{\partial\theta} P_n^m(\cos(\theta))\right)$$

Letting

$$u = P_{n'}^m(\cos(\theta)) \qquad v = P_n^m(\cos(\theta))$$
$$\frac{\partial u}{\partial\theta} = \frac{\partial}{\partial\theta} P_{n'}^m(\cos(\theta)) \quad \frac{\partial v}{\partial\theta} = \frac{\partial}{\partial\theta} P_n^m(\cos(\theta)) \quad VII.14$$

then

$$\int_0^\pi \partial\theta \frac{\partial P_n^m(\cos(\theta))}{\partial\theta} P_{n'}^m(\cos(\theta)) \quad VII.15$$

$= P_{n'}^{m}(\cos(\theta))P_{n}^{m}(\cos(\theta))|_{0}^{\pi} - \int_{0}^{\pi} \partial\theta \frac{\partial P_{n'}^{m}(\cos(\theta))}{\partial\theta} P_{n}^{m}(\cos(\theta)).$

Therefore, VII.13 can be written as

$$\int_{0}^{\pi} \partial\theta \sin(\theta)\, \mathcal{B}_{nn'm}^{\rho\rho'}(\theta) = (-1)^{(\rho'-t')} m\left(P_{n'}^{m}(\cos(\theta))P_{n}^{m}(\cos(\theta))|_{0}^{\pi}\right) + m\left((-1)^{(\rho-t)} - (-1)^{(\rho'-t')}\right) \int_{0}^{\pi} \partial\theta \frac{\partial P_{n'}^{m}(\cos(\theta))}{\partial\theta} P_{n}^{m}(\cos(\theta)) \quad VII.13$$

Then equation VII.3 is

$$\int_{0}^{\pi} \partial\theta \sin(\theta)\, X_{nm}^{\rho-t}(\theta) X_{n'm}^{\rho'-t'}(\theta) = \frac{(-1)^{(\rho-t)(n+m+1)}}{\sqrt{2n(n+1)}} \frac{(-1)^{(\rho'-t')(n'+m+1)}}{\sqrt{2n'(n'+1)}} \begin{pmatrix} n(n+1)\delta_{nn'} \\ +\left((-1)^{((\rho-t)+(\rho'-t'))} - 1\right) \int_{0}^{\pi} \partial\theta\, m^{2} \frac{P_{n}^{m}(\cos(\theta))}{\sin(\theta)} P_{n'}^{m}(\cos(\theta)) \\ +m\left((-1)^{(\rho-t)} - (-1)^{(\rho'-t')}\right) \int_{0}^{\pi} \partial\theta \frac{\partial P_{n'}^{m}(\cos(\theta))}{\partial\theta} P_{n}^{m}(\cos(\theta)) \end{pmatrix} \quad VII.14$$

If $(\rho - t) = (\rho' - t')$ then,

$$\int_{0}^{\pi} \partial\theta \sin(\theta)\, X_{nm}^{\rho-t}(\theta) X_{n'm}^{\rho'-t'}(\theta) = \frac{\delta_{nn'}}{2} \quad VII.15$$

If $m = 0$ then,

$$\int_{0}^{\pi} \partial\theta \sin(\theta)\, X_{nm}^{\rho-t}(\theta) X_{n'm}^{\rho'-t'}(\theta) = \frac{\delta_{nn'}}{2} (-1)^{(\rho-t)(n+1)} (-1)^{(\rho'-t')(n+1)} \quad VII.16$$

## VIII. FAR FIELD FORWARD AND BACKWARD INTENSITY IN THE KERKER BASIS

In the forward direction, the far field polar functions are simplified to

$$\pi_{n}^{|m|}(\beta = 0) = \tau_{n}^{|m|}(\beta = 0) = \frac{1}{2} H_{n} \delta_{|m|1}. \quad VIII.1$$

where, $H_{n} = \frac{1}{\sqrt{2}} \sqrt{n(n+1)(2n+1)}$. Therefore,

$$X_{nm}^{t}(\theta = 0) = \frac{(-1)^{t(n+m+1)}}{\sqrt{2n(n+1)}} \frac{1}{2} H_{n} \delta_{|m|1} (1 + (-1)^{t} m) = \frac{\sqrt{n(n+1)(2n+1)}\delta_{m1}\delta_{t0}}{\sqrt{4n(n+1)}} = \frac{\sqrt{(2n+1)}\delta_{m1}\delta_{t0}}{2} \quad VIII.2$$

In the backward direction, the far field polar functions have the relationship,

$$\pi_{n}^{m}(\pi) = -(-1)^{n} \pi_{n}^{m}(0)$$
$$\tau_{n}^{m}(\pi) = (-1)^{n} \tau_{n}^{m}(0) \quad VIII.3$$

Therefore,

$$X_{nm}^{t}(\theta = \pi) = \frac{(-1)^{t(n+m+1)}}{\sqrt{2n(n+1)}} (-1)^{n} \frac{1}{2} H_{n} \delta_{|m|1} (1 - (-1)^{t} m) = \frac{(-1)^{(n+2)}(-1)^{n}}{\sqrt{2n(n+1)}} H_{n} \delta_{m1} \delta_{t1} = \frac{\sqrt{(2n+1)}\delta_{m1}\delta_{t1}}{2} \quad VIII.4$$

Starting from the azimuthally integrated far field intensity,

$$\int_{0}^{2\pi} r^{2} \partial\phi \langle \boldsymbol{S}_{far} \rangle \cdot \hat{\boldsymbol{e}}_{r} = Re\left\{ \frac{\pi}{\mu\omega k r} \sum_{nn'mp} c_{nmp}^{f} c_{n'mp}^{*f} X_{nm}^{f}(\theta) X_{n'm}^{f}(\theta) + c_{nmp}^{b} c_{n'mp}^{*b} X_{nm}^{b}(\theta) X_{n'm}^{b}(\theta) \right\} = Re\left\{ \frac{\pi}{\mu\omega k} \sum_{tmp} \left\| \sum_{n} c_{nmp}^{t} X_{nm}^{t}(\theta) \right\|^{2} \right\}, \quad VIII.5$$

Therefore,

$$I(\theta = 0^\circ) = Re\left\{\frac{\pi}{\mu\omega k}\sum_p \left\|\sum_n c^f_{n1p}\frac{\sqrt{(2n+1)}}{2}\right\|^2\right\} = \frac{\pi}{2\mu\omega k}\sum_p\left\|\sum_n c^f_{n1p}\, K_n\right\|^2 , \qquad VIII.6$$

where $K_n = \sqrt{(2n+1)}$.

## IX. TOTAL SCATTERED POWER IN THE KERKER BASIS

The total scattered power can be found from integrating the azimuthally integrated power over the polar angles,

$$\int_0^{2\pi} r^2 \sin(\theta)\, \partial\phi\partial\theta\langle \boldsymbol{S}_{far}\rangle \cdot \hat{\boldsymbol{e}}_{\boldsymbol{r}} = Re\left\{\frac{\pi}{\mu\omega kr}\sum_{tnn'mp} c^t_{nmp} c^{*t}_{n'mp}\int_0^\pi \partial\theta \sin(\theta)\, X^t_{nm}(\theta) X^t_{n'm}(\theta)\right\} \qquad IX.1$$

The solution to equation IX.1 involves solving the integral ,

$$\int_0^\pi \partial\theta \sin(\theta)\, X^t_{nm}(\theta) X^t_{n'm}(\theta)$$
$$= \int_0^\pi \partial\theta \sin(\theta)\frac{(-1)^{t((n+n')+2m+2)}}{\sqrt{2n(n+1)}\sqrt{2n'(n'+1)}}\left(\tau^m_n(\theta) + (-1)^t m\pi^m_n(\theta)\right)\left(\tau^m_{n'}(\theta) + (-1)^t m\pi^m_{n'}(\theta)\right) \qquad IX.2$$
$$= \frac{(-1)^{t(n+n')}}{2\sqrt{n(n+1)}\sqrt{n'(n'+1)}}\int_0^\pi \partial\theta \sin(\theta)\begin{pmatrix}\tau^m_n(\theta)\tau^m_{n'}(\theta)\\ +(-1)^t m\pi^m_{n'}(\theta)\tau^m_n(\theta)\\ +(-1)^t m\pi^m_n(\theta)\tau^m_{n'}(\theta)\\ +m^2\pi^m_n(\theta)\pi^m_{n'}(\theta)\end{pmatrix}$$

Using the identities,

$$\int_0^\pi\left(\tau^{|m|}_n(\cos\theta)\tau^{|m|}_{n'}(\cos\theta) + m\pi^{|m|}_n(\cos\theta) m\pi^{|m|}_{n'}(\cos\theta)\right)\sin(\theta)\, d\theta = n(n+1)\delta_{nn'}, \qquad IX.3$$

and

$$\int_0^\pi\left(\tau^{|m|}_n(\cos\theta) m\pi^{|m|}_{n'}(\cos\theta) + \tau^{|m|}_{n'}(\cos\theta) m\pi^{|m|}_n(\cos\theta)\right)\sin(\theta)\, d\theta = 0 \qquad IX.4$$

Equation IX.2 is

$$\int_0^\pi \partial\theta \sin(\theta)\, X^t_{nm}(\theta) X^t_{n'm}(\theta) = \frac{(-1)^{t(n+n')}}{2\sqrt{n(n+1)}\sqrt{n'(n'+1)}} D_{nm}\delta_{nn'} = \frac{1}{2}\delta_{nn'} \qquad IX.5$$

Therefore, equation IX.1 is

$$\int_0^{2\pi} r^2 \sin(\theta)\, \partial\phi\partial\theta\langle \boldsymbol{S}_{far}\rangle \cdot \hat{\boldsymbol{e}}_{\boldsymbol{r}} = Re\left\{\frac{\pi}{2\mu\omega k}\sum_{tnmp}\left\|c^t_{nmp}\right\|^2\right\} \qquad IX.6$$

## X. ILLUSTRATIVE EXAMPLE FROM THE INTRODUCTION

The purpose of this section is to show the mathematical transform between the Mie and Kerker harmonics for the system example proposed in the introduction of the main text. Profiles of the time-averaged far field Poynting vector of each system are shown for illustrative values of $a, b, c$. The main manuscript introduces two systems:

$$\begin{aligned}&System\ 1: \quad -2a\boldsymbol{\Psi}^E_{111} - 2ib\boldsymbol{\Psi}^E_{211} + c\boldsymbol{\Psi}^E_{311} - ic\boldsymbol{\Psi}^M_{310}\\ &System\ 2: \quad -2a\boldsymbol{\Psi}^E_{111} - 2b\boldsymbol{\Psi}^M_{210} + c\boldsymbol{\Psi}^E_{311} - ic\boldsymbol{\Psi}^M_{310},\end{aligned} \qquad X.1$$

where $a, b, c \in \mathbb{Z}^+ \ll \infty$. These two systems can be written in the Kerker basis by transforming the coefficients using the formula

$$\mathrm{c}_{nmp}^{t} = \frac{1}{2}(-1)^{t(n-m-1)}(-i)^{n}\begin{pmatrix} c_{nmp}^{M} \\ +(-1)^{1-t-p} i c_{nm1-p}^{E} \end{pmatrix}. \qquad X.2$$

The transformed coefficients are shown in the table below

| $\mathbf{c}_{nmp}^{M}$ | $\mathbf{c}_{nm1-p}^{E}$ | Kerker Coef. Transform | $\mathbf{c}_{nmp}^{f}$ | $\mathbf{c}_{nmp}^{b}$ |
|---|---|---|---|---|
| $\boldsymbol{c_{110}^{M} = 0}$ | $c_{111}^{E} = -2a$ | $\mathrm{c}_{110}^{t} = \frac{-i}{2}(-1)^{t}(-1)^{1-t} i c_{110}^{E}$ | $\mathrm{c}_{110}^{f} = a$ | $\mathrm{c}_{110}^{b} = a$ |
| $\boldsymbol{c_{210}^{M} = -2b}$ | $c_{211}^{E} = 0$ | $\mathrm{c}_{210}^{t} = \frac{1}{2}(-i)^{2} c_{210}^{M}$ | $\mathrm{c}_{210}^{f} = b$ | $\mathrm{c}_{210}^{b} = b$ |
| $\boldsymbol{c_{210}^{M} = 0}$ | $c_{211}^{E} = -2ib$ | $\mathrm{c}_{210}^{t} = -\frac{i}{2}(-1)^{1-t}(c_{211}^{E})$ | $\mathrm{c}_{210}^{f} = b$ | $\mathrm{c}_{210}^{b} = -b$ |
| $\boldsymbol{c_{310}^{M} = -ic}$ | $c_{311}^{E} = c$ | $\mathrm{c}_{310}^{t} = \frac{1}{2}(-1)^{t}(-i)^{3}(ic)(-1 + (-1)^{1-t})$ | $c_{310}^{f} = c$ | $c_{310}^{b} = 0$ |

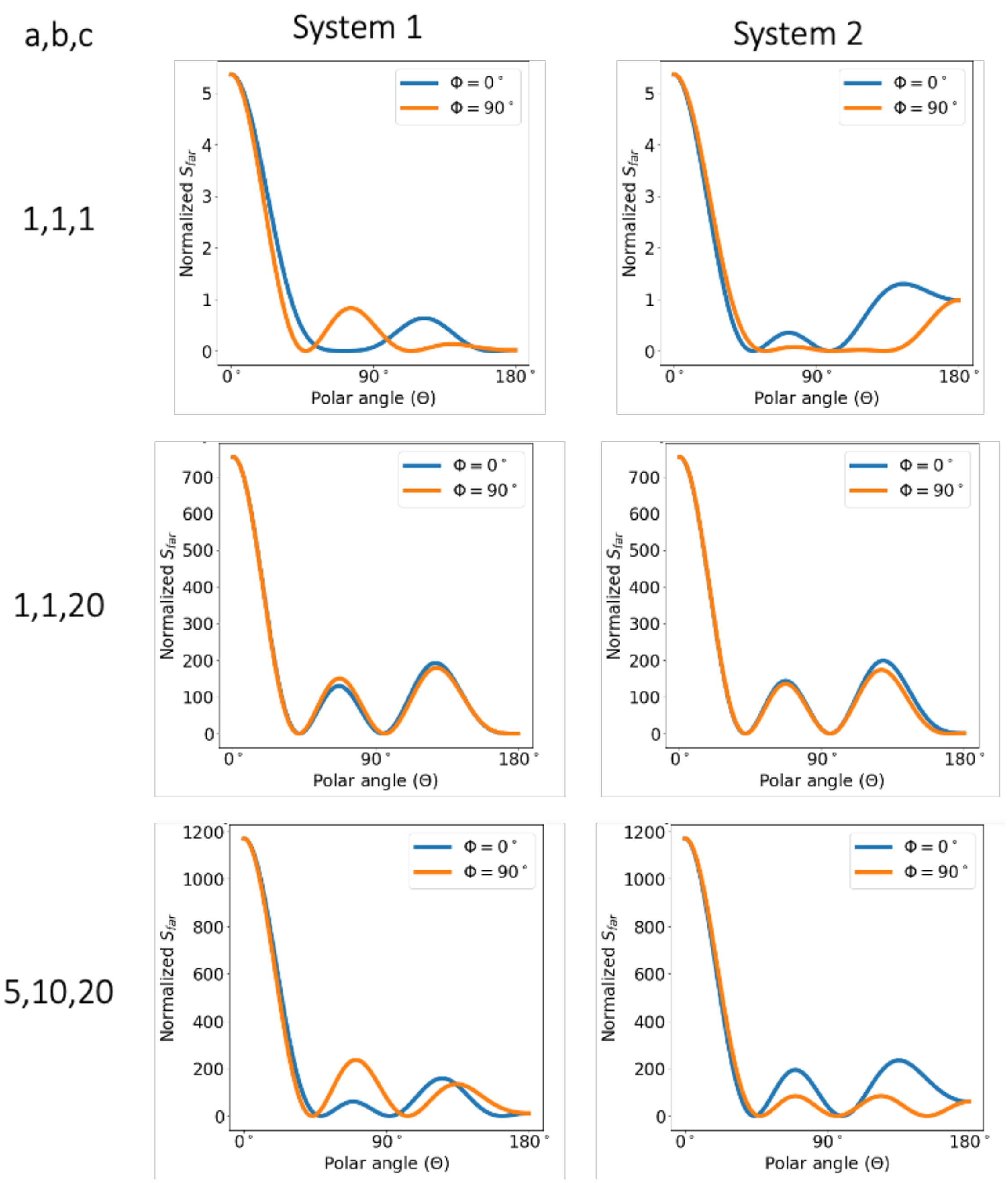

Figure S1: Time-averaged Poynting vector of system 1 (middle column) and system 2 (right column) proposed in equation X.1 for different values of a, b, and c (left column). The blue line is for $\phi = \pi/2$ the orange line is for $\phi = 0$. The time-averaged Poynting vector is normalized to $2\mu\omega k r^2\langle \boldsymbol{S}_{far}\rangle = (\|A(\phi,\theta)\|^2 + \|B(\phi,\theta)\|^2)\hat{\boldsymbol{e}}_r$, where $A(\phi,\theta)$ and $B(\phi,\theta)$ come from equation 9 in the main text.

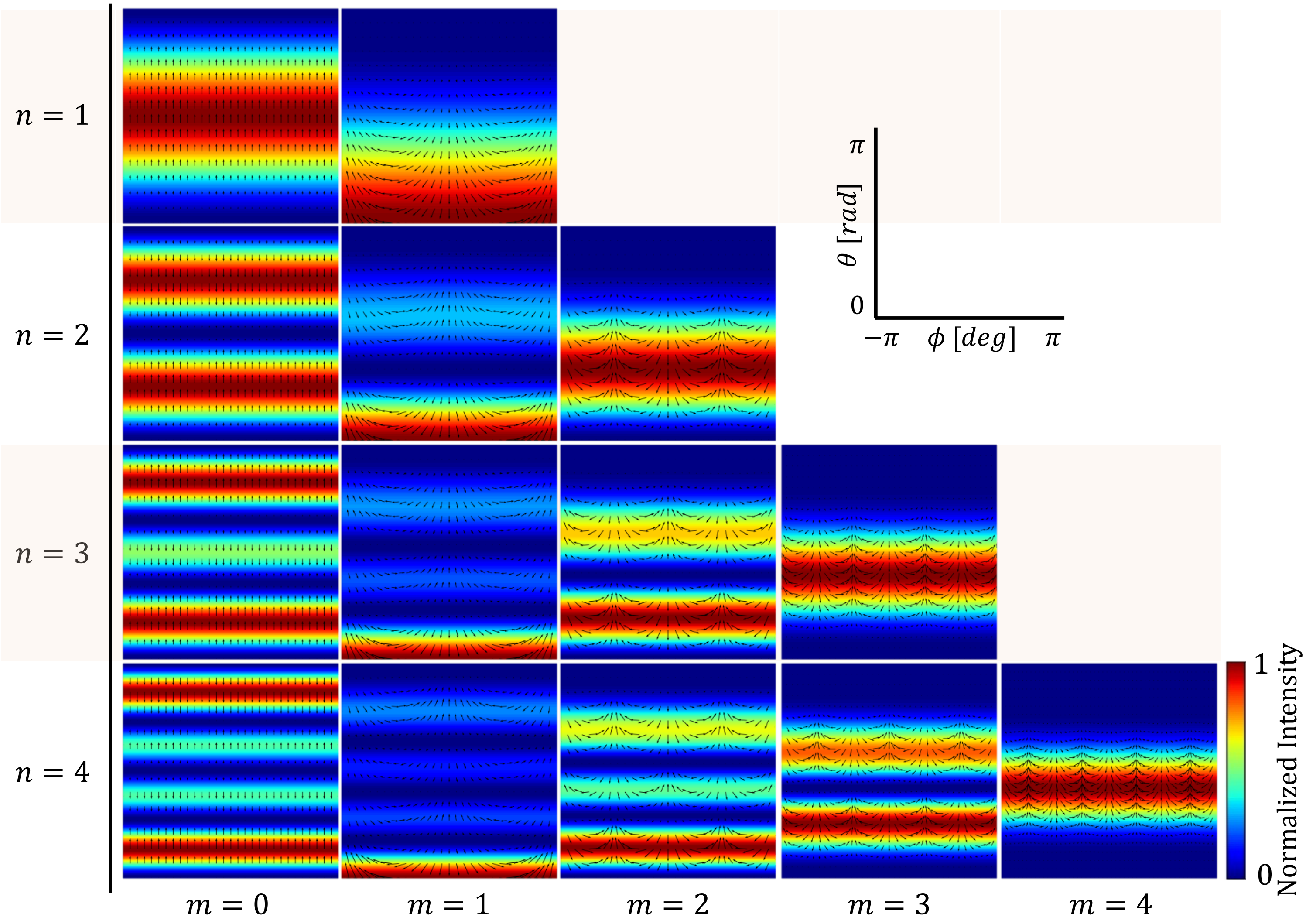

Figure S2: Color coded time-averaged intensity of the far field forward Kerker harmonics with vector polarization overlaid as black arrows. In each figure, the x-axis is the azimuthal angle and the y-axis is the polar angle.